\documentclass[12pt,superscriptaddress]{article}
\pdfoutput=1
\usepackage{jheppub}
\usepackage{graphicx}  
\usepackage{dcolumn}   
\usepackage{bm}        
\usepackage{amssymb}   
\usepackage{amsmath}
\usepackage{slashed} 

\newcommand{\beq}{\begin{equation}}
\newcommand{\eeq}[1]{\label{#1}\end{equation}}
\def\beqa{\begin{eqnarray}}
\def\eeqa#1{\label{#1}\end{eqnarray}}
\newcommand{\eeqn}{\end{equation}}
\newcommand{\CR}{\notag \\}
\newcommand{\leqn}[1]{(\ref{#1})}
\newcommand{\hhref}[1]{\href{http://arxiv.org/abs/#1}{arXiv:#1}}

\def\stacksymbols #1#2#3#4{\def\theguybelow{#2}
    \def\vp{\lower#3pt}
    \def\sp{\baselineskip0pt\lineskip#4pt}
    \mathrel{\mathpalette\intermediary#1}}

\def\intermediary#1#2{\vp\vbox{\sp
     \everycr={}\tabskip0pt
     \halign{$\mathsurround0pt#1\hfil##\hfil$\crcr#2\crcr
              \theguybelow\crcr}}}

\def\gsim{\stacksymbols{>}{\sim}{2.5}{.2}}
\def\lsim{\stacksymbols{<}{\sim}{2.5}{.2}}
\begin{document}

\title{Higgs Couplings and Naturalness in $\lambda$-SUSY}

\author{Marco Farina$^1$,}	

\author{Maxim Perelstein$^1$,}
	
\author{Bibhushan Shakya$^{1,2}$}
\affiliation[1]{Laboratory for Elementary Particle Physics,
	     Cornell University, Ithaca, NY 14853, USA}	
\affiliation[2]{Michigan Center for Theoretical Physics,\\ Department of Physics, University of Michigan, Ann Arbor, MI 48109, USA}
	
\emailAdd{mf627@cornell.edu}
\emailAdd{mp325@cornell.edu}
\emailAdd{bshakya@umich.edu}

\abstract{We study Higgs boson couplings in the large-$\lambda$ version of the Next-to-Minimal Supersymmetric Standard Model, known as $\lambda$-SUSY. We find that the predicted deviations from the Standard Model (SM) in these couplings are inversely correlated with the amount of fine-tuning needed to accommodate a 126 GeV Higgs. In the most natural regions of parameter space, the 126 GeV Higgs has large admixtures of both the SM-singlet and the non-SM Higgs doublet scalars, and such regions are already ruled out by the LHC. Future improvements in the Higgs coupling measurements will either discover deviations from the SM, or put further stress on naturalness in $\lambda$-SUSY.}

\maketitle

\newpage

\section{Introduction}

Over the past two years, the LHC experiments reported overwhelming evidence for the existence of a new particle, with a mass of approximately 126 GeV. The new particle's properties appear roughly consistent with the Higgs boson, predicted by the Standard Model (SM) and incorporated in many of its popular extensions, such as supersymmetric (SUSY) models. However, in the Minimal Supersymmetric Standard Model (MSSM), the mass of the observed particle, well in excess of the upper bound on the tree-level Higgs mass inherent in the structure of the model, leads to well-known tension with naturalness. (Actually, the observation of the 126 GeV Higgs just exacerbated the already serious fine-tuning issue faced by the MSSM since the LEP-2 experiment in the late 1990's~\cite{Barbieri:2000gf}.) This tension led to a revival of interest in non-minimal realizations of SUSY at the weak scale. Arguably the simplest among such extended constructions is the Next-to-Minimal Supersymmetric Standard Model (NMSSM), which will be the focus of this paper. In the NMSSM, additional contributions to the tree-level Higgs mass, not present in the MSSM, are generated, allowing to accommodate a 126 GeV Higgs with significantly less fine-tuning.

The particle content of the NMSSM consists of all fields of the MSSM, plus a chiral superfield $\hat{S}$, uncharged under any of the SM gauge groups. (For a recent comprehensive review, see Ref.~\cite{Ellwanger:2009dp}.) In the simplest version of the theory, which we study here, the Higgs sector superpotential has the form
\beq
W = \lambda \hat{S} \hat{H}_u \cdot \hat{H}_d + \kappa \hat{S}^3\,,
\eeq{dub-ya}
where $\hat{H}_u$ and $\hat{H}_d$ are the usual Higgs superfields, $\lambda$ and $\kappa$ are dimensionless coefficients, and we defined the $SU(2)$ index contraction $A\cdot B \equiv \epsilon_{ab} A^a B^b$. This is the most general superpotential consistent with a ${\cal Z}_3$ discrete symmetry under which each of the three superfields is charged. Note that the $\mu$-term of the MSSM is forbidden by this symmetry; instead, an effective $\mu$-term is generated when the scalar component of the field $\hat{S}$ gets a vacuum expectation value (vev). This term, in turn, drives electroweak symmetry breaking (EWSB). The tree-level F-term scalar potential contains mass terms for the Higgs fields, proportional to $\lambda^2$; it is these terms, absent in the MSSM, that lift the Higgs mass and reduce fine-tuning.

Numerical values of $\lambda$ and $\kappa$ are free parameters. Traditionally, studies of the NMSSM focused on the region where $\lambda\lsim 0.7$. (Here, and throughout the paper, all numerical values of parameters will refer to their {\it weak-scale} values, unless explicitly specified otherwise.) In this region, the model remains perturbative up to the grand unification (GUT) scale, of order $10^{16}$ GeV. However, this requirement limits the size of the F-term contribution to the Higgs mass, and the issue of naturalness for a 126 GeV Higgs can be addressed only partially. To reduce fine-tuning further, consider the variation of the model in which one does {\it not} require perturbative grand unification. In this scenario, $\lambda$ is allowed to hit a Landau pole below the GUT scale, so that larger values of weak-scale $\lambda$ are allowed. (Nevertheless, models with precision gauge coupling unification can still be constructed~\cite{Hardy:2012ef}.) Imposing the requirement that the Landau pole does not occur below 10 TeV, which would very likely lead to conflict with precision electroweak tests of the SM, yields a requirement $\lambda \lsim 2.0$. The NMSSM with $0.7 \lsim \lambda \lsim 2.0$ has been dubbed $\lambda${\it -SUSY}~\cite{Barbieri:2006bg}. This model can easily accommodate a 126 GeV Higgs, with no need for a significant top-loop contribution to the quartic. Moreover, it has another advantage: The sensitivity of the weak scale to the stop mass scale is reduced by a factor of $\sim(g/\lambda)^2$, compared to the MSSM~\cite{Barbieri:2006bg,Hall:2011aa,Perelstein:2012qg}. (Here $g$ is the SM weak gauge coupling.) This means that the lower bound on the stop mass imposed by the LHC direct searches (currently about 700 GeV, assuming a small LSP mass) has milder implications on fine-tuning in this model than in the MSSM or the NMSSM with $\lambda\lsim 0.7$. Motivated by these arguments, we will focus on the $\lambda$-SUSY regime of the NMSSM in this paper.

While the sensitivity to the stop mass is reduced at large $\lambda$, an additional fine-tuning among tree-level parameters of $\lambda$-SUSY is necessary to accommodate the 126 GeV Higgs mass, as pointed out in~\cite{Agashe:2012zq,Gherghetta:2012gb}. In this paper, we will show that the required fine-tuning is further increased when the LHC constraints on the Higgs {\it couplings} are taken into account. In $\lambda$-SUSY, the structure of the 126 GeV ``Higgs boson" is quite complex: in general, it can be a mixture of three gauge eigenstates, two $SU(2)$ doublets and one singlet. For large values of $\lambda$, such mixing is in fact necessary to obtain a 126 GeV Higgs~\cite{Hall:2011aa,Agashe:2012zq}. This structure results in deviations of the Higgs couplings from SM predictions. We will systematically explore these deviations,\footnote{Constraints on the NMSSM from the LHC Higgs couplings measurements have been previously studied in Refs.~\cite{King:2012tr,Gupta:2012fy,Cheung:2013bn,Cheng:2013fma,Barbieri:2013hxa}; however, questions of naturalness were not considered in those papers. Constraints on the NMSSM from the early $h\to\gamma\gamma$ data, and their fine-tuning implications, were considered in~\cite{Cao:2012yn}.} and conclude that generically, their magnitude is {\it inversely correlated} with the amount of fine-tuning required to accommodate the observed Higgs mass.  In the most natural regions of the parameter space, the 126 GeV Higgs has large admixtures of both the weak singlet and the non-SM weak doublet Higgs states. As a result, such natural regions are already ruled out by the LHC Higgs rate measurements. Future experiments at the LHC, including a luminosity upgrade, and possibly at a next-generation $e^+e^-$ ``Higgs factory", such as the proposed International Linear Collider (ILC), will improve the coupling measurement precision from the current 20-30\% to $\sim1$\% in many cases~\cite{Peskin:2012we,Baer:2013cma}. As precision improves, either a deviation from the SM will be discovered, or $\lambda$-SUSY will become progressively more fine-tuned.

Interestingly, we find one possible exception to these trends, a very small ``anomalous" region of the parameter space where relatively low fine-tuning ($\sim 1/10$) can be achieved. However, while this is intriguing, our tree-level analysis is not sufficiently accurate to establish the stability of the EWSB vacuum, as well as consistency with the LHC Higgs data, in this region. Further work is required to address this issue.

The paper is organized as follows. We briefly describe the model, and discuss the theoretical and experimental constraints determining the viable region of its parameter space, in Sec.~\ref{sec:setup}. Sec.~\ref{sec:FTmeasure} describes the quantitative measure of fine-tuning used in our analysis. The results of the analysis are presented in Sec.~\ref{sec:tb1}, which discusses the case $\tan\beta=1$, which can be treated almost completely analytically, and in Sec.~\ref{sec:scans}, which presents the results of our numerical exploration of the full parameter space. The scatter plots in that section illustrate the main conclusions of the paper. The main conclusions are summarized in Sec.~\ref{sec:conc}.

\section{The Model}
\label{sec:setup}

We work in the setup of the ``scale-invariant" NMSSM, with the superpotential given in Eq.~\leqn{dub-ya}, and follow the notation of Refs.~\cite{Ellwanger:2009dp,Agashe:2012zq}. The scalar potential for the Higgs fields $H_u$, $H_d$ and $S$ is given by the sum of the usual F- and D-term contributions, and the soft SUSY breaking terms:
\beq
V_{\rm soft} =  m_u^2 |H_u|^2 + m_d^2 |H_d|^2 + m_S^2 |S|^2 + \left( \lambda A_\lambda S H_u\cdot H_d + \frac{1}{3} \kappa A_\kappa S^3 +~{\rm h.c.}\right).
\eeq{Vsoft}
In all, the Higgs sector Lagrangian contains 7 free parameters:
\beq
p_i = \{\lambda, \kappa, m_u^2, m_d^2, m_S^2, A_\lambda, A_\kappa\}.
\eeq{pis}
We will assume all parameters to be real; there is neither explicit nor spontaneous CP violation in the Higgs sector of this model~\cite{Romao:1986jy}.
In the realistic vacuum ({\it i.e.} a stable vacuum exhibiting EWSB) the neutral components of $H_u$ and $H_d$, as well as the singlet field $S$, get vacuum expectation values (vevs): $\langle H_u\rangle = v_u$,  $\langle H_d\rangle = v_d$, $\langle S\rangle = s$, where $v\equiv \sqrt{v_u^2+v_d^2} \approx 174$ GeV.  These vevs are obtained from the minimization equations of the Higgs potential
\beqa
\nonumber E_1\equiv m_{H_u}^2+\mu^2+\lambda^2v_d^2+\frac{g^2}{2}(v_u^2-v_d^2)-\frac{v_d}{v_u}\mu(A_\lambda+\kappa s)=0\,, \nonumber\\
 E_2\equiv m_{H_d}^2+\mu^2+\lambda^2v_u^2+\frac{g^2}{2}(v_d^2-v_u^2)-\frac{v_u}{v_d}\mu(A_\lambda+\kappa s)=0 \,,\nonumber\\
E_3\equiv  m_{S}^2+\frac{\kappa}{\lambda}A_\kappa \mu+2\frac{\kappa^2}{\lambda^2}\mu^2+\lambda^2 (v_u^2+v_d^2) -2\lambda\kappa v_u v_d-\lambda^2v_uv_d\frac{A_\lambda}{\mu}=0\,,
\eeqa{E1E3}
where $m_Z^2=g^2v^2, \mu=\lambda s$. We defined $g^2=(g_1^2+g_2^2)/2 \approx 0.52$, where $g_1$ and $g_2$ are the SM $U(1)_Y$ and $SU(2)_L$ couplings respectively.

\begin{figure}[tb]
\begin{center}
\centerline {
\includegraphics[width=5.5in]{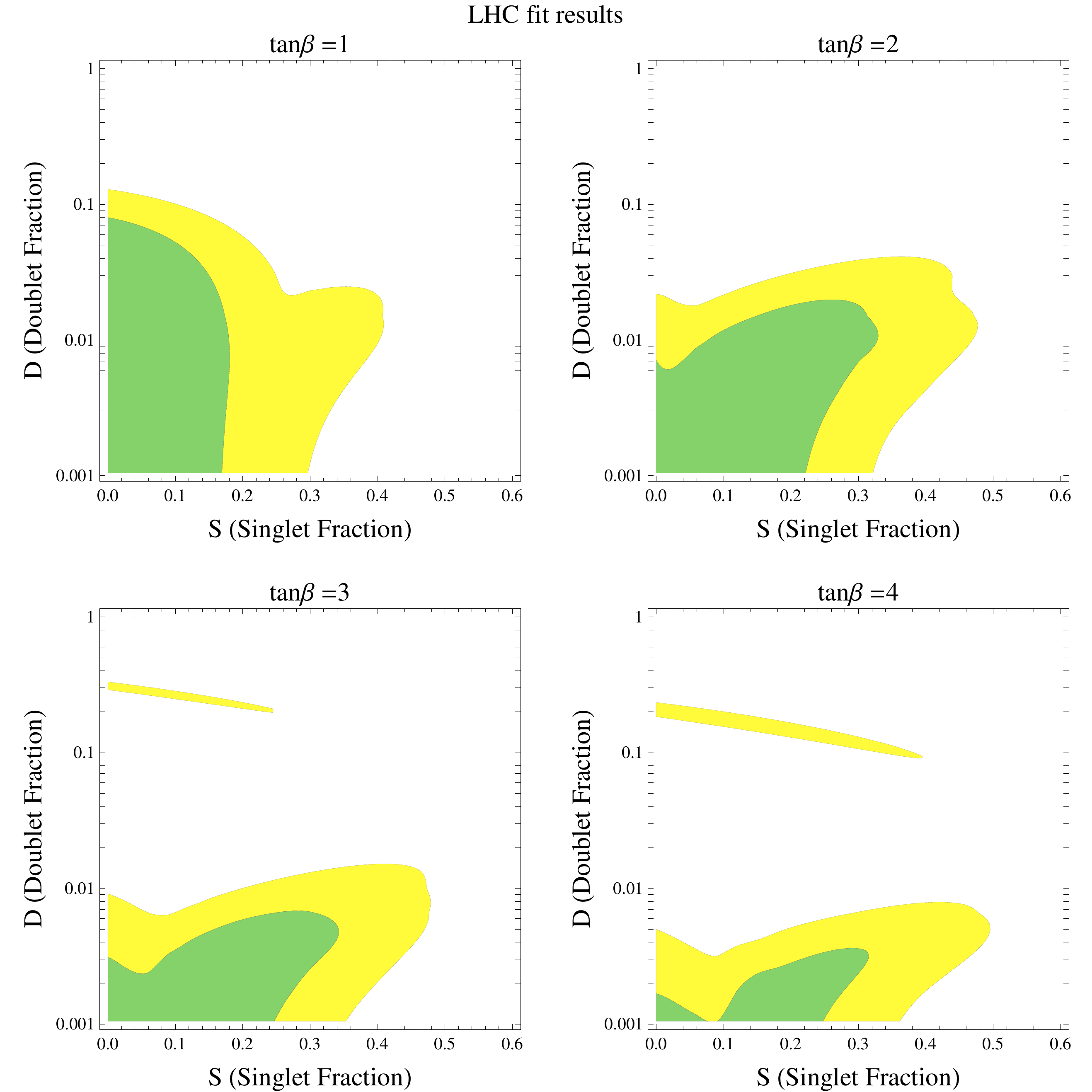}
}
\caption{LHC bounds on the non-SM doublet admixture $D$ and singlet admixture $S$ in the 126 GeV Higgs particle. Green and yellow regions correspond to $68\%$ and $95\%$ C.L. allowed by Higgs data fit. }
\label{fig:bounds}
\end{center}
\end{figure}

Expanding around the vacuum yields the physical Higgs fields: three CP-even and two CP-odd electrically neutral states, plus one charged Higgs. (An additional 2 neutral and 1 charged degrees of freedom are the Goldstone bosons absorbed by the SM gauge bosons.) For the CP-even fields, it is convenient to work in the basis $(h_v^0, H_v^0, h_s^0)$, defined by
\beqa
H_u^0 &=& v_u + \frac{1}{\sqrt{2}}\left( \sin\beta \, h_v^0 + \cos\beta \, H_v^0\right), \CR
H_d^0 &=& v_d + \frac{1}{\sqrt{2}}\left( \cos\beta \, h_v^0 - \sin\beta \, H_v^0\right), \CR
S &=& s + \frac{1}{\sqrt{2}} h_s^0\,,
\eeqa{basis}
where $\tan\beta = v_u/v_d$.
The advantage of this particular basis is that, of the three fields, only $h_v^0$ has tree-level couplings to the SM $W/Z$ bosons. Since these couplings have been shown experimentally to exist for the 126 GeV particle, with strengths roughly consistent with the SM expectations, it is clear that the 126 GeV boson has to at least have a significant component along $h_v^0$. In other words, if we write the 126 GeV mass eigenstate as
\beq
h^0_{{\rm 126}} = \alpha_h h_v^0 + \alpha_H H_v^0 + \alpha_s h_s^0\,,
\eeq{126}
and define the ``non-SM doublet admixture" $D=|\alpha_H|^2$ and the ``singlet admixture" $S=|\alpha_s|^2$, the LHC data puts constraints on $D$ and $S$. These constraints, based on our fit to the rates reported by the LHC and TeVatron experiments \cite{ATLASbb,ATLAStautau,ATLASww,ATLASzz,ATLASgaga,ATLASgaga1212,CMSbb,CMStautau,CMSww,CMSzz,CMSgaga,Tevatron}, are shown in Fig.~\ref{fig:bounds}. The constraints depend on the value of $\tan\beta$, which enters into the couplings of $h_v^0$ and $H_v^0$ to SM fermions. We show the constraints for $\tan\beta = 1 \ldots 4$; this is the range most interesting in $\lambda$-SUSY, as will be discussed later. The value of $D$ allowed by the fits varies between $1\%$ and $10\%$ depending on $\tan \beta$, with the exception of a narrow strip at larger $D$ which is allowed at a 2$\sigma$ level. (In this strip, the Higgs couplings to SM fermions happen to have their SM values up to an overall phase of $-1$.) The maximum allowed mixing with the singlet is always around $30-50\%$. In summary, the data essentially points to the $h-H$ decoupling limit, while still allowing large singlet mixing.

In the $(h_v^0, H_v^0, h_s^0)$ basis, the CP-even Higgs mass$^2$ matrix is given by
{\small
\beq
{\cal M}^2 = \left( \begin{tabular}{ccc}
$\lambda^2 v^2 \sin^2 2\beta + m_Z^2 \cos^2 2\beta$ & $\frac{1}{2} (\lambda^2v^2-m_Z^2) \sin4\beta$ & $2\lambda v \left[ \mu - \left( \kappa s + \frac{1}{2} A_\lambda \right) \sin2\beta\right]$ \\

$\cdot$ & ($m_Z^2-\lambda^2 v^2)\sin^2 2\beta +\frac{2B\mu}{\sin2\beta}$&$-2\lambda v \left( \kappa s + \frac{1}{2} A_\lambda\right)\cos 2\beta$ \\

$\cdot$ & $\cdot$ &$\kappa s (4\kappa s + A_\kappa) + \frac{v^2}{2s} A_\lambda \lambda \sin 2\beta$ \end{tabular} \right)\,.
\eeq{3by3}
}
Here we used the potential minimization conditions~\leqn{E1E3} to trade the parameters $m_u^2$, $m_d^2$ and $m_S^2$ for $m_Z$, $\tan\beta$, and $s$, and defined $\mu=\lambda s$ and $B=A_\lambda + \kappa s$. We will require that the lowest eigenvalue of this matrix is $m_h^2=(126$~GeV$)^2$. One of the model parameters can then be eliminated in favor of $m_h$; a convenient choice is to eliminate $A_\kappa$, since it enters linearly into the characteristic equation for $m_h^2$, and is thus unambiguously fixed.

Before proceeding, we note that the mass matrix above, and all other formulas used in the bulk of the analysis of this paper, are tree-level only.  Loop corrections can be important~\cite{Ellwanger:2005fh,Degrassi:2009yq,Staub:2010ty}. In particular, top and stop loops can give a substantial contribution to the CP-even Higgs masses. In the gauge basis, {\it i.e.} before the rotation of Eq.~\leqn{basis}, the one-loop correction to the mass of the up-type Higgs boson has the form
\beq
\delta m^2 \approx \frac{3y_t^2 m_t^2}{4\pi^2} \,\ln \frac{m_{\tilde{t}_1}m_{\tilde{t}_2}}{m_t^2}\,,
\eeq{mhcorr}
where $m_{\tilde{t}_i}$ are the masses of the two stops, and $m_{\tilde{t}}\gg m_t$ is assumed.
Upon rotation, this term contributes to the upper-left $2\times 2$ block of the mass matrix~\leqn{3by3}. We will briefly consider the effect of this correction at the end of the paper, and show that our qualitative conclusions do not change for reasonable values of stop masses. We will not consider corrections due to loops of the Higgs-sector fields themselves. While possibly significant due to large values of $\lambda$ of interest, these loops depend sensitively on the masses of the Higgs-sector superpartners, which are at present very poorly constrained by the data. We leave a detailed analysis of the loop corrections for future work.

\begin{figure}[tb]
\begin{center}
\centerline {
\includegraphics[width=2.9in]{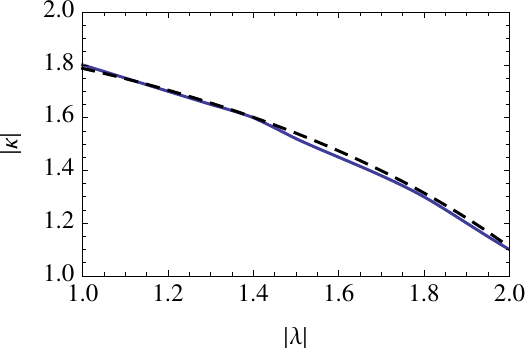}
}
\caption{The allowed region in the $\lambda-\kappa$ plane. The boundary, denoted by the blue curve, is obtained by requiring no Landau pole below 10 TeV, as estimated through RG running of the couplings at two loops \cite{King:1995vk, Masip:1998jc, Ellwanger:2009dp}. The black dashed curve shows the analytic approximation to the boundary, Eq.~\leqn{kappaboundaryapprox}.}
\label{fig:kappacutoff}
\end{center}
\end{figure}

To summarize, the Higgs sector of our model (at tree level) is completely described by five parameters:
\beq
\{ \lambda, \kappa, \tan\beta, s, A_\lambda\}.
\eeq{oompa}
The region of interest in this space is determined by the following considerations:

\begin{itemize}

\item $1.0\lsim \lambda \lsim 2.0$: As already explained in the Introduction, this is the interesting region for $\lambda$-SUSY. It should be emphasized that the reduction of sensitivity of the weak scale to the stop mass, an important advantage of the model, is maximized at larger $\lambda$, scaling as $\sim(\lambda/g)^2$.

\item $\kappa$: While $|\kappa|\lsim 0.65$ is required in the NMSSM to avoid a Landau pole below the GUT scale, this consideration is no longer relevant in $\lambda$-SUSY, and considerably larger values of $\kappa$ can be realized. For a given value of $\lambda$, the maximum possible weak-scale value of $\kappa$ can be estimated by requiring that $\kappa$ does not hit a Landau pole below 10 TeV. We use two-loop renormalization group (RG) equations \cite{King:1995vk, Masip:1998jc, Ellwanger:2009dp} to perform this estimate. The allowed region in the $\lambda-|\kappa|$ plane is shown in Fig. \ref{fig:kappacutoff}. The boundary can be conveniently approximated as
\beq
0.17\,\lambda^2+0.26\,\kappa^2=1.
\eeq{kappaboundaryapprox}

\item $1\leq \tan\beta \leq 4$: Precision electroweak constraints disfavor $\tan\beta\gsim 4$~\cite{Gherghetta:2012gb, Barbieri:2006bg, Franceschini:2010qz}. In addition, the reduction of sensitivity of the weak scale to the stop mass is lost at large $\tan\beta$, motivating $\tan\beta\sim 1$.

\item $s$, $A_\lambda$: These two dimensionful parameters can in principle take any value. However it is clear that taking them well above the weak scale would result in increased fine-tuning. We will consider $|s|, |A_\lambda|\leq 10$ TeV.

\end{itemize}

Furthermore, only a subset of this parameter space describes realistic models. To identify this subset, we impose the following constraints:

\begin{itemize}

\item Lightest CP-even Higgs is the 126 GeV state, and there are no CP-even tachyons. Another interesting case in which the lightest scalar is mostly singlet with mass below 126 GeV has been recently explored~\cite{Agashe:2012zq,Belanger:2012tt,Kang:2013rj,Badziak:2013bda,Barbieri:2013nka,Cao:2013gba}. However this only occurs for low values of $\lambda$, outside of the region of interest in this paper, and we will not consider this possibility.

\item No tachyonic CP-odd or charged states (the CP-odd and charged Higgs masses are given in~\cite{Ellwanger:2009dp,Agashe:2012zq}). Note that the LHC searches for the CP-odd Higgs \cite{CMS:gya} are not relevant for our analysis, since in the MSSM, they do not place a bound below $\tan\beta<5$, while in the NMSSM these bounds will be further weakened by the reduction of the CP-odd Higgs production cross sections due to mixing with the CP-odd singlet.

\item Doublet and singlet admixtures in the 126 GeV Higgs consistent with the fit to the LHC Higgs data, at $95$\%~c.l. (see Fig.~\ref{fig:bounds}).

\item Chargino mass above LEP-2 bound $m_{\chi^+}\geq 94$ GeV~\cite{Beringer:1900zz}. Note that the chargino masses also depend on the wino mass parameter $M_2$, which is unconstrained in our analysis. However the eigenvalues cannot get lower than $\sqrt{\mu^2+ g_2^2 v^2 \sin 2\beta/2}$, which corresponds to the choice $M_2=-\mu$. We therefore impose a conservative lower bound $|\mu|\geq \sqrt{(94~{\rm GeV})^2- g_2^2 v^2 \sin 2\beta/2} $.

\item The neutralino contribution to the invisible width of the $Z$ boson is limited to be less than one standard deviation of the measured neutrino contribution: $\Gamma_{Z\rightarrow\chi\chi}\,\textless \,4.2\,$MeV if $m_\chi \,\textless\, m_Z/2$~\cite{PDG,Hooper:2002nq}. This constraint depends sensitively on the bino mass parameter $M_1$, since the bino component of the neutralino does not couple to the $Z$, reducing the contribution to invisible width. In our numerical analysis, we scan over $M_1$ to take this into account.

\item Stability of the realistic EWSB vacuum with respect to tunneling into unrealistic vacua~\cite{Kanehata:2011ei,Kobayashi:2012xv}.

\end{itemize}

The last constraint deserves a more detailed discussion. The scalar potential of the NMSSM has several local extrema. One of them is the ``realistic" vacuum, with non-zero values of $v_u$, $v_d$ and $s$, and the observed values of $m_Z$ and $m_h$. Since we require that no tachyons are present in this vacuum, it is a local minimum of the potential. However, it is not guaranteed that it is a global minimum; some of the other vacua may have lower energies, in which case the realistic vacuum would be unstable with respect to tunneling into these lower-lying ones. The vacuum energy in the realistic vacuum is\footnote{Eq.~\leqn{Vr} corrects typos in some of the coefficients of Eq.~(28) of Ref.~\cite{Agashe:2012zq}.}
\beqa
V_r &=& - \lambda^2 \frac{m_Z^4 \sin^2 2\beta}{4g^4} - \frac{m_Z^4 \cos^2 2\beta}{4g^2} - \frac{\kappa^2\mu^4}{\lambda^4} - \frac{\kappa A_\kappa \mu^3}{3\lambda^3} \CR & & -\frac{1}{2} \mu v^2 \left[ 2\mu - \sin 2\beta \left( A_\lambda + \frac{2\kappa}{\lambda} \mu \right) \right] \, .
\eeqa{Vr}
To evaluate the stability of this vacuum, we compute the energies of the other, ``unrealistic" vacua, $V_{ui}$, by numerically solving the potential minimization equations~\leqn{E1E3}. If $V_r$ is found to be close to the lowest-lying $V_{ui}$, the situation is somewhat ambiguous, for two reasons. First, our analysis is tree-level only, and loop corrections may affect the relative depths of the vacua and reverse the hierarchy. Second, even if $V_r$ lies above one of the unrealistic minima, it may still be metastable, potentially with lifetime longer than the age of the universe. A detailed analysis of these issues is beyond the scope of this work; here, we take a conservative approach and accept points with $V_r > V_{ui}$ as long as they are relatively close to each other. The points are rejected only if
\beq
V_r-V_{u,min}>\delta \,\,(|V_r| + |V_{u,min}|) ~~~\text{and}~~~|V_r|>\delta \,v^4,
\eeq{delta_def}
where $V_{u,min}$ is the lowest of the unrealistic potentials, and $\delta$ is a numerical constant (roughly, a fractional difference between $V_r$ and $V_{u,min}$). In the following sections, we assume $\delta=0.2$; we checked that varying this parameter within a reasonable range does not affect our conclusions. The second condition, $|V_r|>\delta \,v^4$, is necessary because the loop correction to $V_r$ is not expected to be $\ll V_r$ if $V_r$ is accidentally close to zero. (Indeed, an additive constant can always be added to the potential.) Instead, the expected size of the loop correction is roughly $\delta V_r \sim L v^4$, where $L$ is the loop factor. If $V_r$ and $V_u$ are both close to zero, their order can be reversed by loop corrections even if the first condition in~\leqn{delta_def} is satisfied. We conservatively accept such points, but tag them to indicate that the stability of the realistic vacuum is uncertain (see section~\ref{sec:scans}).

\section{Quantifying Fine-Tuning}
\label{sec:FTmeasure}

As a quantitative measure of naturalness, we use the sensitivity of the (SM-like) Higgs mass to the underlying Lagrangian parameters:\footnote{Note that the underlying parameters used to measure fine-tuning are defined at the {\it weak} scale, not a high scale as is customary in the MSSM literature. A purely weak-scale measure of fine-tuning in the MSSM has been advocated in~\cite{Perelstein:2007nx,Baer:2012up,Baer:2012mv}, and its relation to the traditional measures was discussed in~\cite{Baer:2013gva}. In $\lambda$-SUSY, no perturbative extrapolation to high scales is possible, leaving weak-scale tuning as the only available measure.}
\beq
\Delta = \max_i \left| \frac{\partial \log m_h^2 }{\partial \log p_i}\right|\,,
\eeq{FTdef}
where $i$ runs over the 7 Lagrangian parameters in~\leqn{pis}. Note that $m_h^2$ is obtained by diagonalizing the matrix~\leqn{3by3}, in which the Lagrangian parameters $m_u^2$, $m_d^2$ and $m_S^2$ have been swapped for $m_Z$, $\tan\beta$, and $s$. In this form, the chain rule should be used to compute the derivatives  in Eq.~\leqn{FTdef}: for example,
\beqa
\frac{\partial \log m_h^2}{\partial \log A_\lambda} &=& \frac{\partial \log m_h^2}{\partial \log A_\lambda}|_{m_Z, t_\beta, s} + \left( \frac{\partial \log m_h^2}{\partial m_Z}|_{t_\beta, s} \right) \left(\frac{\partial m_Z}{\partial \log A_\lambda} \right)+ \CR & & \left(\frac{\partial \log m_h^2}{\partial t_\beta}|_{m_Z, s} \right)\left(\frac{\partial t_\beta}{\partial \log A_\lambda} \right)+\left( \frac{\partial \log m_h^2}{\partial s}|_{m_Z, t_\beta} \right)\left(\frac{\partial s}{\partial \log A_\lambda}\right)\,,
\eeqa{chainrule}
where $t_\beta\equiv \tan\beta$. Note that this fine-tuning measure implicitly includes the sensitivity of the weak scale $v$ to the Lagrangian parameters, via derivatives such as $\frac{\partial m_Z}{\partial \log A_\lambda}$ in the above expression.

A convenient way to compute the derivatives of $m_Z$, $t_\beta$, and $s$ with respect to Lagrangian parameters is to use the constraint that the minimization conditions~\leqn{E1E3} must continue to hold under variations of the input parameters ~\cite{ftmeasure,Perelstein:2012qg}. This yields
\beq
\delta E_j=\sum_i\frac{\partial E_j}{\partial \xi_i}\delta \xi_i+\frac{\partial E_j}{\partial m_Z^2}\delta m_Z^2+\frac{\partial E_j}{\partial t_\beta}\delta t_\beta+\frac{\partial E_j}{\partial \mu}\delta\mu=0\,,
\eeq{pv}
for $j=1\ldots 3$, where $i$ runs over the fundamental parameters listed in Eq.~\leqn{pis}. These three equations can be solved for $\delta m_Z^2,~\delta\tan\beta,$ and $\delta s$. Defining
\begin{equation}
\frac{\partial E_j}{\partial \xi_i}=P_{ij},~~~\frac{\partial E_j}{\partial m_Z^2}=Z_j,~~~ \frac{\partial E_j}{\partial \tan\beta}=T_j,~~~ \frac{\partial E_j}{\partial \mu}=M_j\,,
\end{equation}
we obtain, for example, the derivatives of $m_Z$ with respect to Lagrangian parameters:
\begin{equation}
 \frac{\partial\,m_Z^2}{\partial\,\text{log}\,\xi_i} = \xi_i\frac{\delta m_Z^2}{\delta \xi_i}=-\xi_i\frac{\sum_{jkl}\epsilon^{jkl}~P_{ij}T_kM_l}{\sum_{jkl}\epsilon^{jkl}~Z_jT_kM_l}\,.
\label{delta-NMSSM}
\end{equation}
The derivatives of $t_\beta$ and $s$ are obtained through permutations of $M, T, Z$.

It is important to remember that the measure used here is only sensitive to fine-tunings in the tree-level potential. There may be additional sources of fine-tuning at loop level, for example large loops in the top sector if stops are heavy. This tuning would not show up in $\Delta$. The correct interpretation of $\Delta$ is as the {\it minimal} amount of fine-tuning possible for a given parameter point, regardless of the stop masses and other parameters entering only at loop level.

\section{A Simple Limit: $\tan\beta=1$}
\label{sec:tb1}

We first consider the limit $\tan\beta=1$. In this limit, the ${\mathcal{M}}^2_{12},{\mathcal{M}}^2_{23}$ entries of the Higgs mass matrix vanish, see Eq.~\leqn{3by3}. The heavier, non-SM-like Higgs doublet completely decouples, as preferred by the LHC data. Another strong motivation for considering this limit is its simplicity: the Higgs sector effectively consists of two fields, $h_v^0$ and $h_s^0$, and almost all relevant calculations are analytically tractable. The insights obtained in this analysis carry over to the more complicated case of $\tan\beta\not=1$, which has to be treated mostly numerically and will be considered in the next section.

The first simple observation is that the doublet diagonal mass$^2$ term, ${\cal M}_{11}^2 = \lambda^2 v^2$,
is {\it larger} than $m_h^2=(126~{\rm GeV})^2$, throughout the interesting parameter space of $\lambda$-SUSY~\cite{Hall:2011aa}. This means that an admixture of a singlet in the 126 GeV state is not just generic, but is in fact {\it required} in this model. Furthermore, obtaining a 126 GeV Higgs requires fine-tuning among the elements of the mass matrix, especially at large $\lambda$~\cite{Agashe:2012zq,Gherghetta:2012gb}. This can be most easily seen by considering the limit ${\cal M}_{11}^2 \gg m_h^2$. In this limit,
\beq
m_{h}^2 \approx {\mathcal{M}}^2_{11}-\frac{{\mathcal{M}}^4_{13}}{{\mathcal{M}}^2_{33}}\,.
\eeq{mhapp}
The two terms on the right-hand side have to cancel with precision of order $m_{h}^2/{\mathcal{M}}^2_{11}=m_h^2/( \lambda^2 v^2)$. For example, for $\lambda =2.0$, this corresponds to roughly 15\% fine-tuning. However, such an estimate constitutes just a starting point. The cancellation in Eq.~\leqn{mhapp} can in fact be natural if the two terms are correlated by the underlying theory. On the other hand, additional fine-tuning may be required in order to get the required values of the ${\cal M}^2$ matrix elements from the potential. In order to address these issues, fine-tuning needs to be measured with respect to Lagrangian parameters, as explained in Section~\ref{sec:FTmeasure}. In addition, this simple estimate does not take into account various constraints, discussed in Section~\ref{sec:setup}, which can increase the fine tuning by ruling out the naively most natural parts of the parameter space. As we will see, this situation is in fact generic, so that including the constraints is crucial for understanding the amount of fine-tuning required.

\begin{figure}[htb]
\begin{center}
\includegraphics[width=0.78\linewidth]{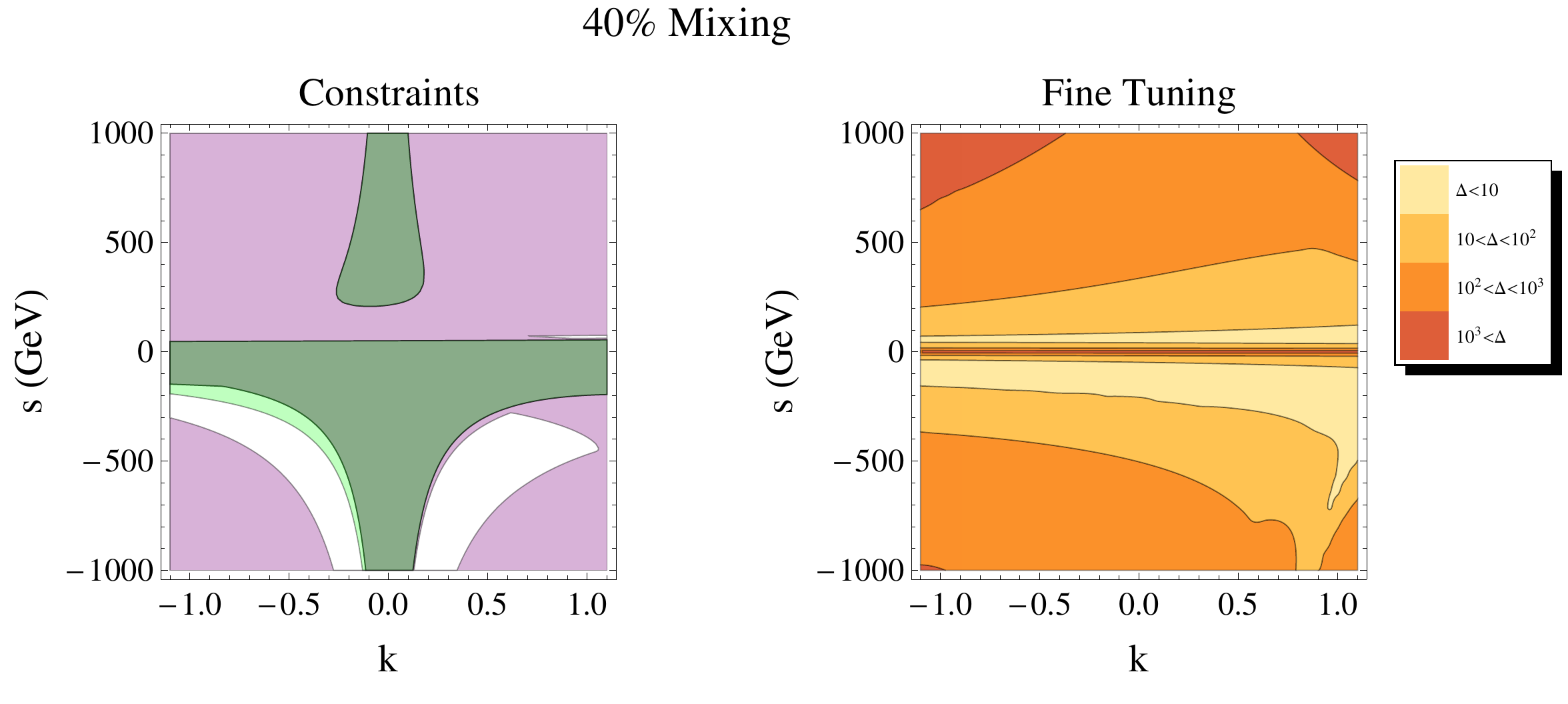}
\includegraphics[width=0.78\linewidth]{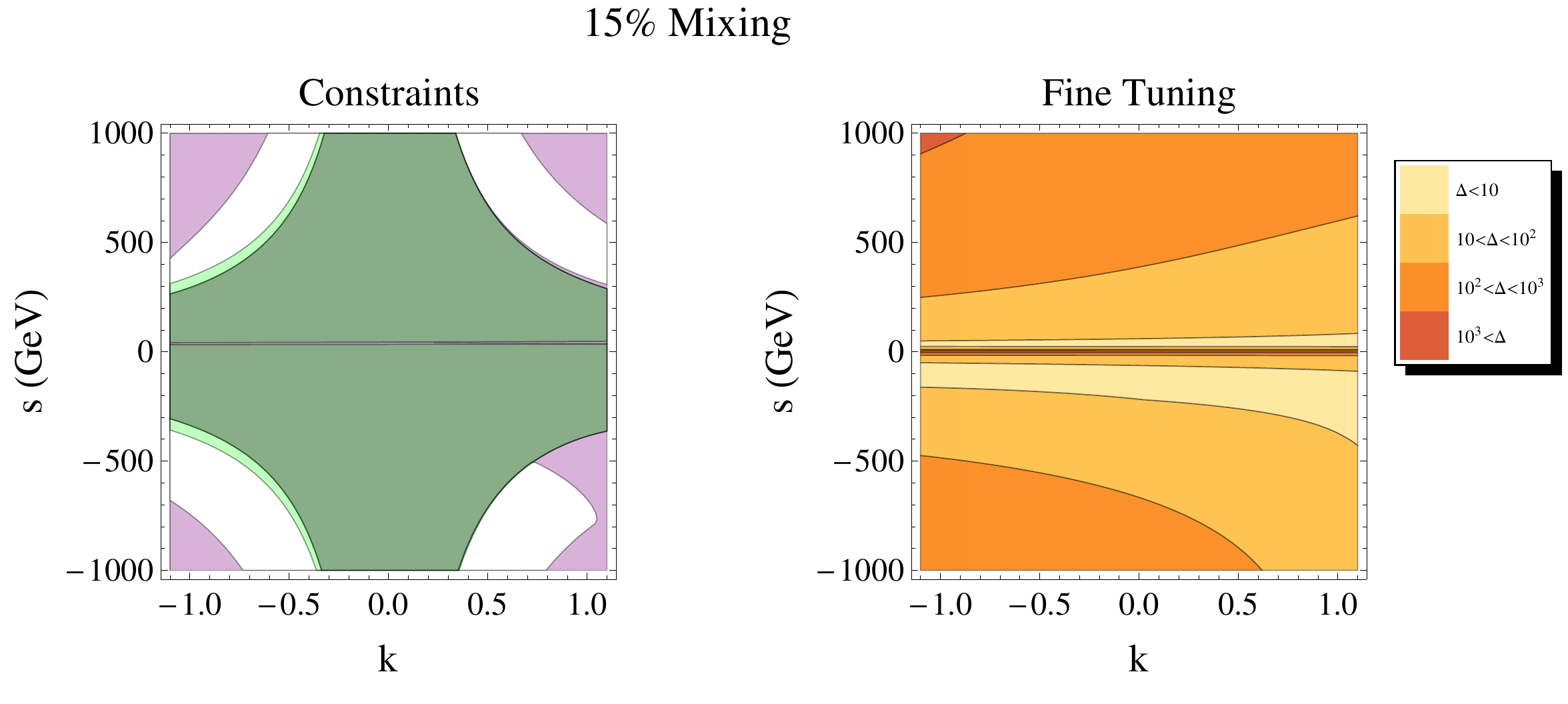}
\includegraphics[width=0.78\linewidth]{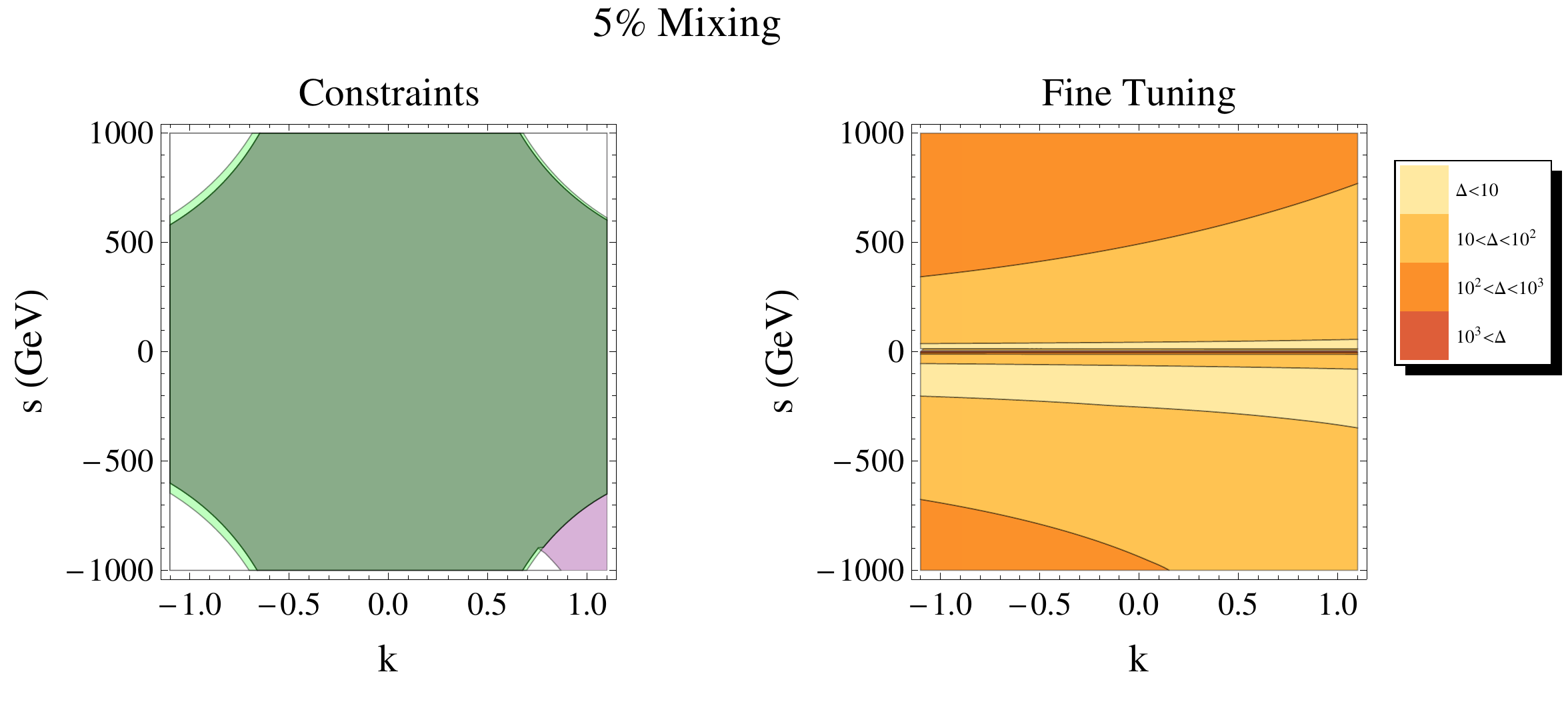}
\end{center}
\caption{Constraints (left) and fine-tuning (right) in the $\kappa-s$ plane, for $\lambda=2$, $\tan\beta=1$ and different values of the singlet fraction $S$: from top to bottom, $S=40\%$, $15\%$ and $5\%$. On the constraints plots, allowed regions are shown in white; regions excluded due to instability of the realistic EWSB minimum are shown in purple; while regions excluded due to presence of tachyonic scalar states and/or experimental constraints are shown in green (for details, see section~\ref{sec:setup}). On the fine-tuning plots, lighter colors represent less tuned regions.}
\label{fig:kVSs}
\end{figure}

With $\tan\beta=1$, the theory is described by the remaining four parameters in Eq.~\leqn{oompa}. The main focus of our analysis is on understanding the correlation between the singlet fraction $S$ of the 126 GeV Higgs and fine-tuning. The singlet fraction is given by
\beq
S = |\sin\phi|^2,
\eeq{Sphi}
where $\phi$ is the mixing angle between the doublet and the singlet. In terms of the fundamental model parameters,
\beq
\phi = \arctan  \frac{\lambda^2 v^2 - m_h^2}{2\lambda v \left( (\lambda-\kappa)s -\frac{1}{2} A_\lambda\right) }.
\eeq{Dexpr}
This can be used to eliminate one of the model parameters in favor of $S$. We choose to eliminate $A_\lambda$.\footnote{For any given $\{\lambda,~\kappa,~s,~S\}$, solving Eqs.~\leqn{Sphi} and~\leqn{Dexpr} results in two possible values of $A_\lambda$, each of which is in turn associated with one $A_k$. On the other hand, the model is invariant under $\{s,~A_\lambda,~A_k\} \rightarrow \{-s,~-A_\lambda,~-A_k\}$, so that only half of the $s$ plane needs to be considered to find all physically distinct solutions. In Fig.~\ref{fig:kVSs}, we plot one of the $\{A_\lambda,~A_k\}$ solutions in the lower half-plane ($s<0$), and the other solution in the upper half-plane ($s>0$).} To analyze the behavior of fine-tuning as a function of $S$, we fix $\lambda$ and plot the fine-tuning contours, as well as constraints, in the $\kappa-s$ plane, for several values of $S$. For example, Fig.~\ref{fig:kVSs} shows a series of plots for $\lambda=2.0$.

The main conclusion is that {\it the minimal required fine-tuning increases with decreasing singlet fraction in the 126 GeV Higgs}. In other words, the most natural regions of the parameter space have a large singlet admixture in the 126 GeV state, and are already in tension with the LHC data. Further improvements of the Higgs rate measurements will either yield a deviation from the SM, or increase the amount of fine-tuning required in $\lambda$-SUSY.

This behavior can be qualitatively understood from two observations. First, from Eq.~\leqn{Dexpr}, it follows that, once $m_h$ and $\lambda$ are fixed, a decrease in the mixing angle $\phi$ can only be achieved by raising the dimensionful parameters $s$ and/or $A_\lambda$. This introduces a hierarchy between these terms and the doublet vev $v$ and therefore leads to fine-tuning. Second, this tension is further increased when constraints on the parameter space are taken into account. Fig.~\ref{fig:kVSs} shows that as $S$ is decreased, the regions allowed by the constraints shift towards larger values of the singlet vev $|s|$, increasing the hierarchy of scales and therefore fine-tuning. It turns out that the most important constraints for understanding the observed behavior are the requirement of the positive CP-odd Higgs mass$^2$, and the stability  of the realistic minimum. The first of these constraints can be approximated as
\beq
\kappa^2 s^2+\left(1-\frac{1}{2 S}\right) \lambda^2 (v^2-m_h^2) \gtrsim 0 \,.
\eeq{Constra}
This formula was obtained analytically by expanding the CP-odd mass matrix at large $s$ and small $\kappa$, and it provides a very good approximation to the exact constraint curves plotted in Fig.~\ref{fig:kVSs}. Using this formula, the correlation observed earlier is easy to understand: for large mixing, the constraint can be satisfied with $s\sim v$, while for small mixing, a hierarchy $s\gg v$ is required, leading to fine-tuning.

\begin{figure}[t!]
\begin{center}
\includegraphics[width=0.7\linewidth]{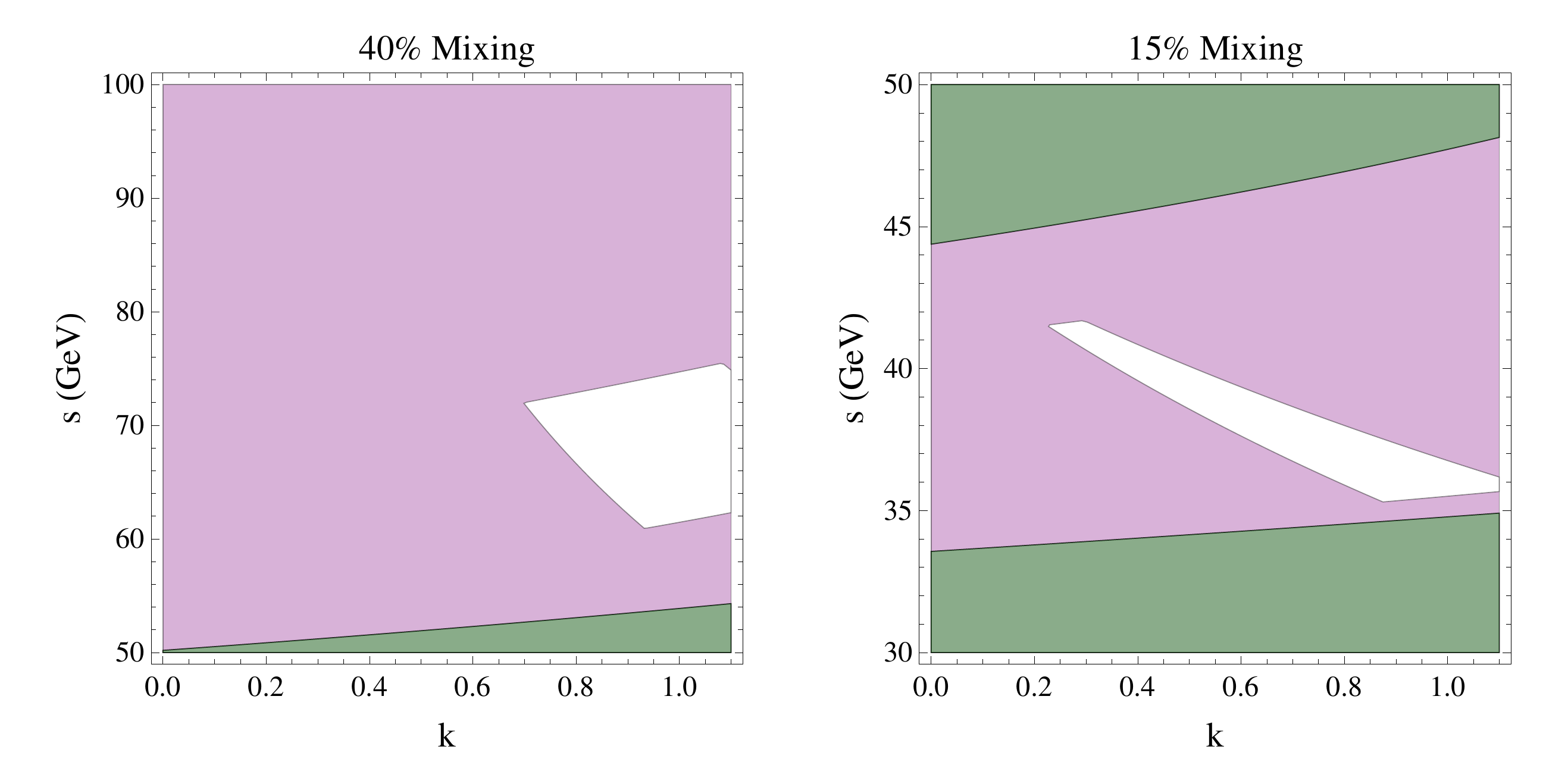}
\end{center}
\caption{The ``anomalous" allowed region in the $\kappa-s$ plane, for $\tan\beta=1$ and two representative values of the singlet fraction. The same color code as in Fig.~\ref{fig:kVSs} is used. No anomalous region was found for 5\% mixing.}
\label{fig:zoom}
\end{figure}

In addition to the sizable ``bulk" allowed regions clearly visible in Fig.~\ref{fig:kVSs}, an additional narrow strip of parameter space is allowed for small $s$, of order 50 GeV, and positive $\kappa$. This ``anomalous" region is difficult to see in Fig.~\ref{fig:kVSs} and is shown in Fig.~\ref{fig:zoom}, which shows a zoom into the appropriate part of parameter space. If indeed viable, this anomalous region would be the most attractive part of the model parameter space from the point of view of naturalness. However, its viability is far from certain. In this region, the energy of the realistic vacuum $V_r$ is accidentally close to zero: $V_r\ll v^4$. For any choice of parameters, the theory has an unrealistic vacuum with $v_u=v_d=s=0$, and the energy of that vacuum is always $V_u=0$. In the anomalous region, our conservative criterion, Eq.~\leqn{delta_def}, indicates that the realistic vacuum could be stable, but the realistic and unrealistic vacua are sufficiently close in energy that their order may well be reversed by loop corrections. An analysis of the full one-loop potential is required to clarify the situation. In addition, the anomalous region is characterized by low ($\sim 100$ GeV) values of $\mu$, and therefore light charginos. Given the large values of $\lambda$ we are interested in, the loops of these particles can have a significant effect on the Higgs branching ratios~\cite{SchmidtHoberg:2012yy,Choi:2012he}, which were not taken into account in our fits. We defer a detailed analysis of the viability of this anomalous region to future work.

\section{Numerical Analysis}
\label{sec:scans}

For general $\tan\beta$, the non-SM-like Higgs doublet $H_v^0$ does not decouple, and the full system of the three CP-even Higgs fields needs to be considered. In this situation, we use a numerical scan of the parameter space to study the correlation between the singlet fraction $S$ and the degree of fine-tuning. We find that the correlations found in the $\tan\beta=1$ case of the previous section still apply.

To generate points for the numerical scans, we first fix $\lambda=2.0$, and choose the other four parameters listed in Eq.~\leqn{oompa} randomly, within the boundaries specified in Section~\ref{sec:setup}. (The slightly asymmetric treatment of $\lambda$ and the other parameters is chosen for ease of comparison with the $\tan\beta=1$ limit.) The points were assumed to be distributed linearly in $\kappa$ and $\tan\beta$ and log-linearly in the dimensional parameters, $s$ and $A_\lambda$. While these choices provide a comprehensive coverage of parameter space, they are not physically motivated; as a result, variations in the relative density of points on the scatter plots below have no physical significance. The only robust and physically relevant features are the {\it boundaries} of the populated and unpopulated regions.

\begin{figure}[t!]
\centering
\begin{tabular}{cc}
\includegraphics[width=3in,height=2in]{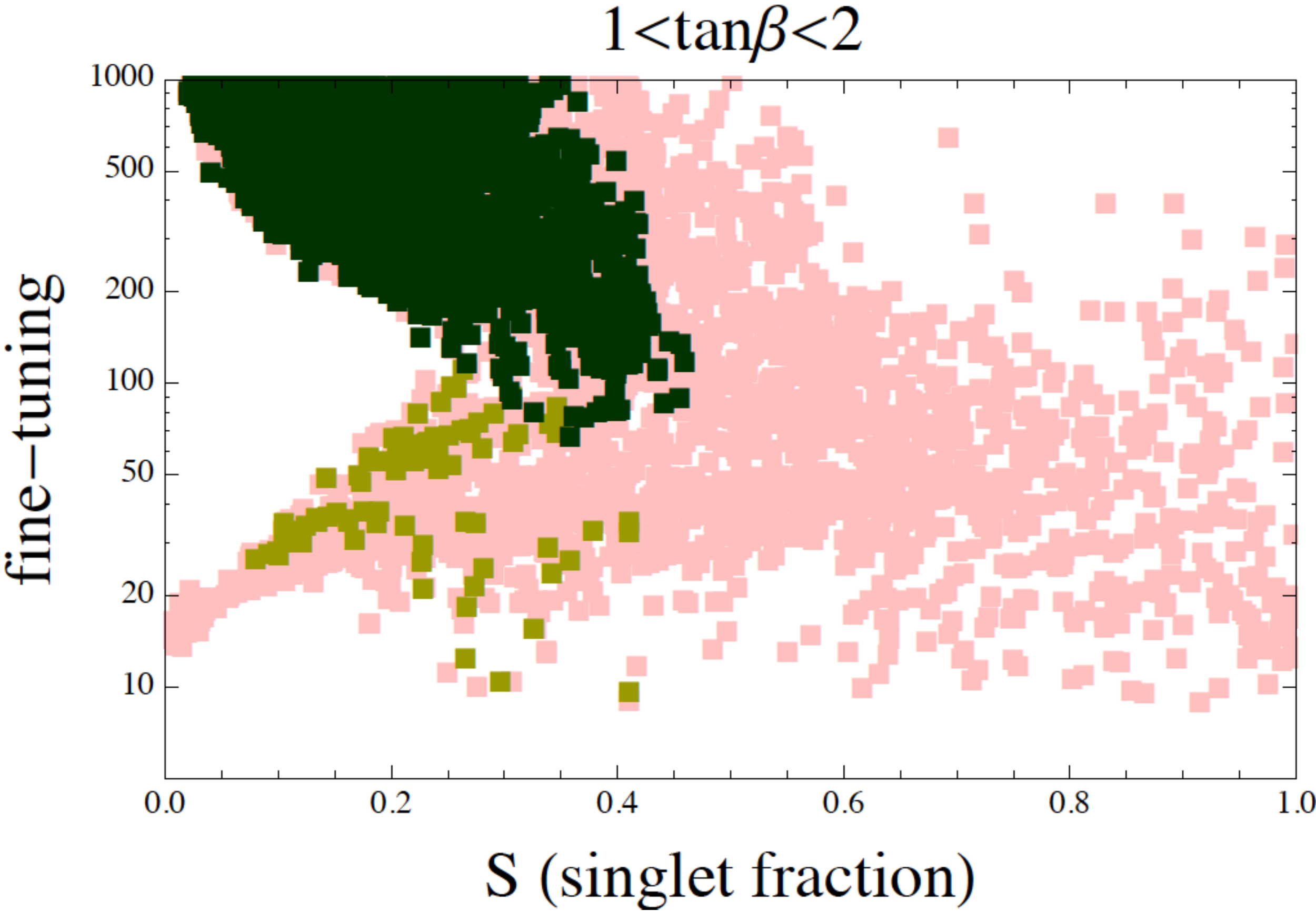}&
\includegraphics[width=3in,height=2in]{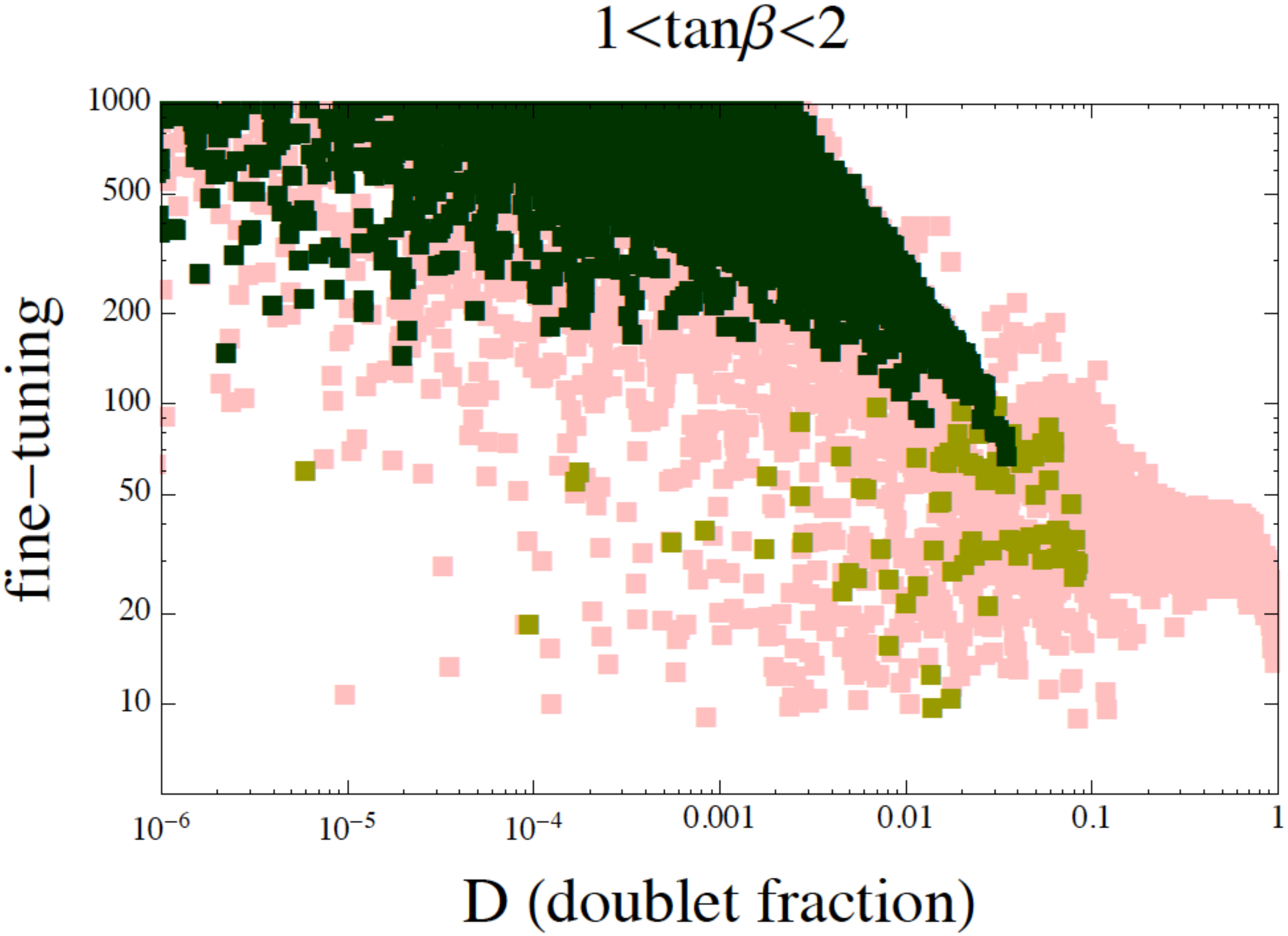}\\
\\
\includegraphics[width=3in,height=2in]{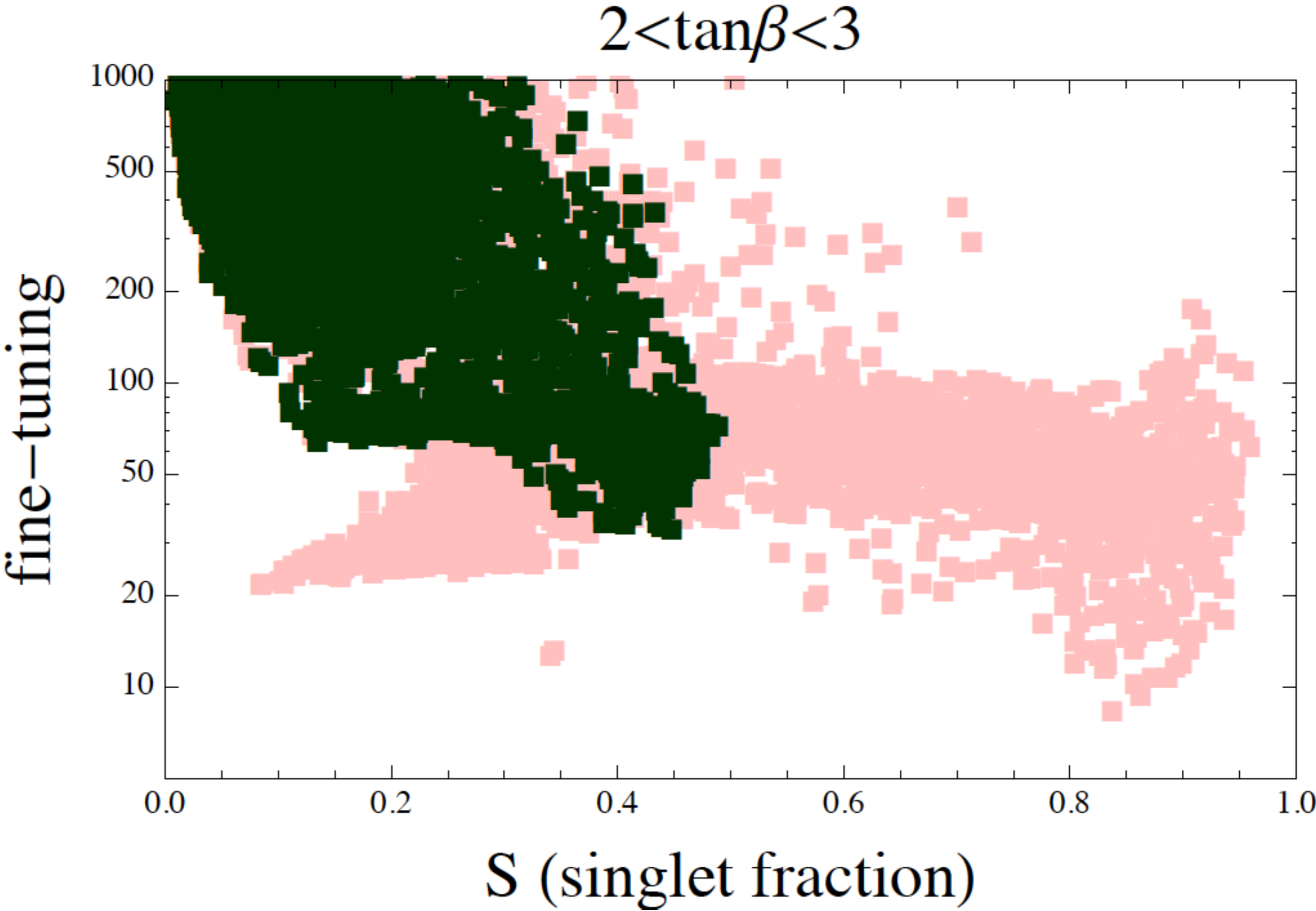}&
\includegraphics[width=3in,height=2in]{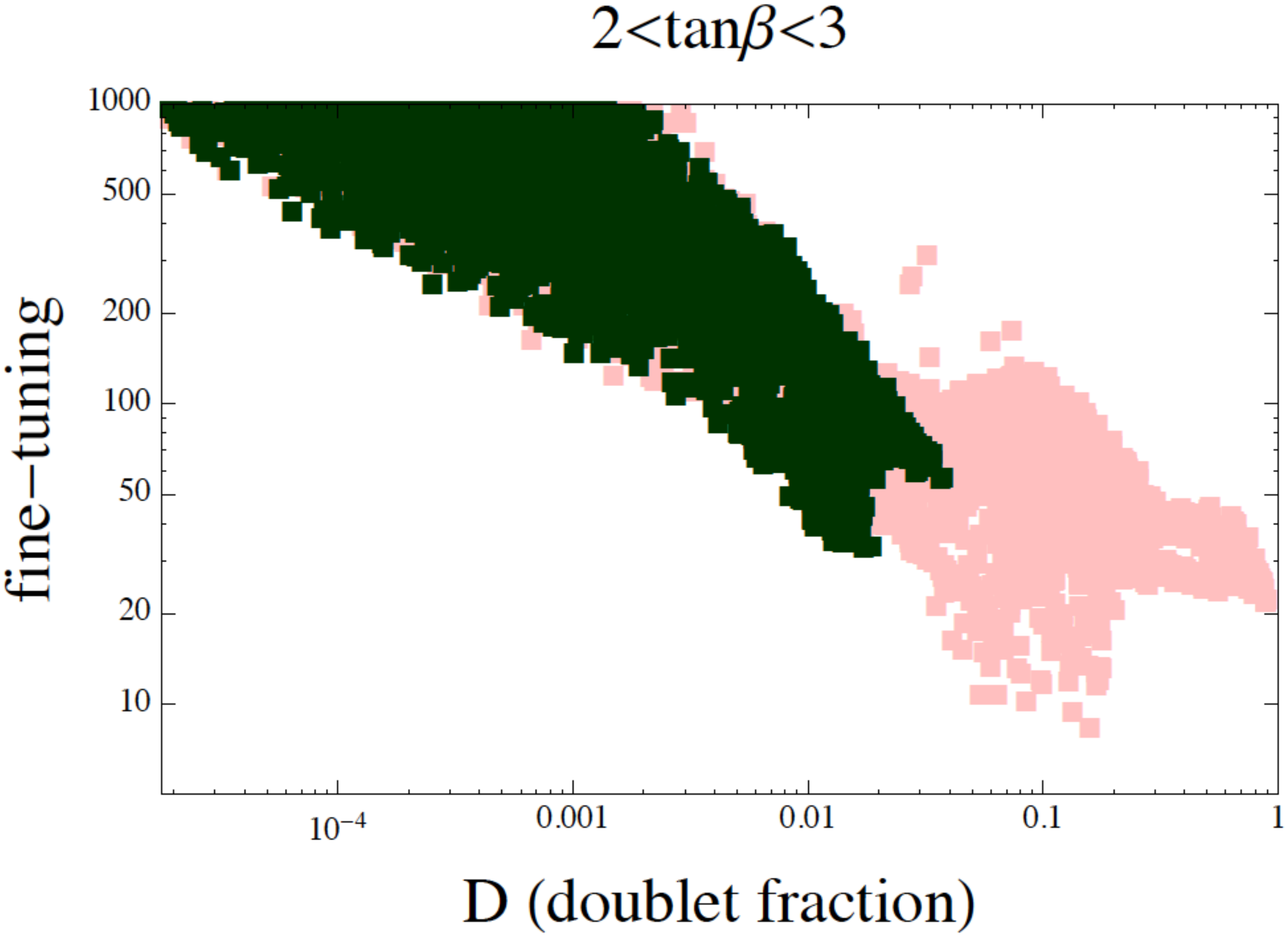}\\
\\
\includegraphics[width=3in,height=2in]{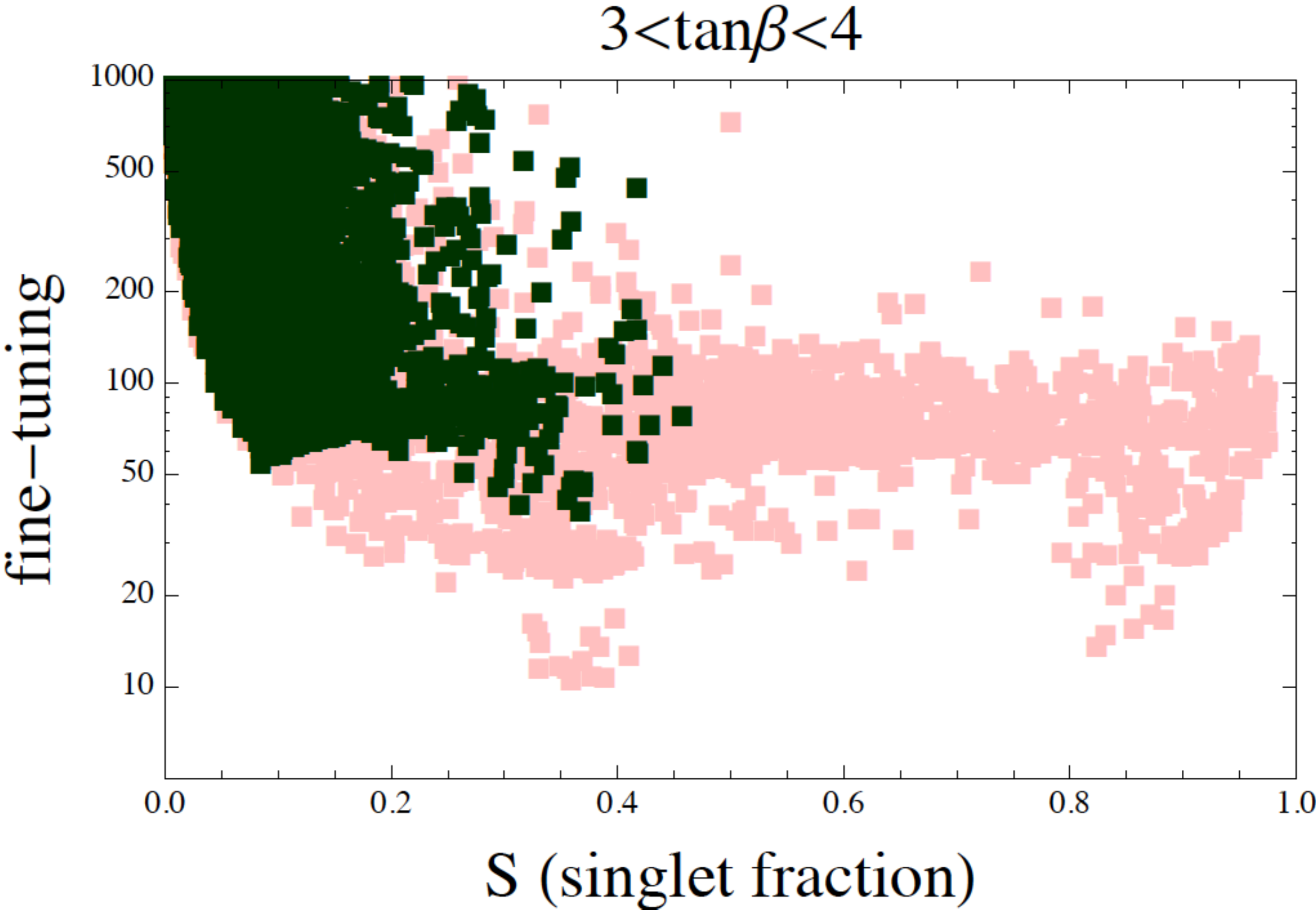}&
\includegraphics[width=3in,height=2in]{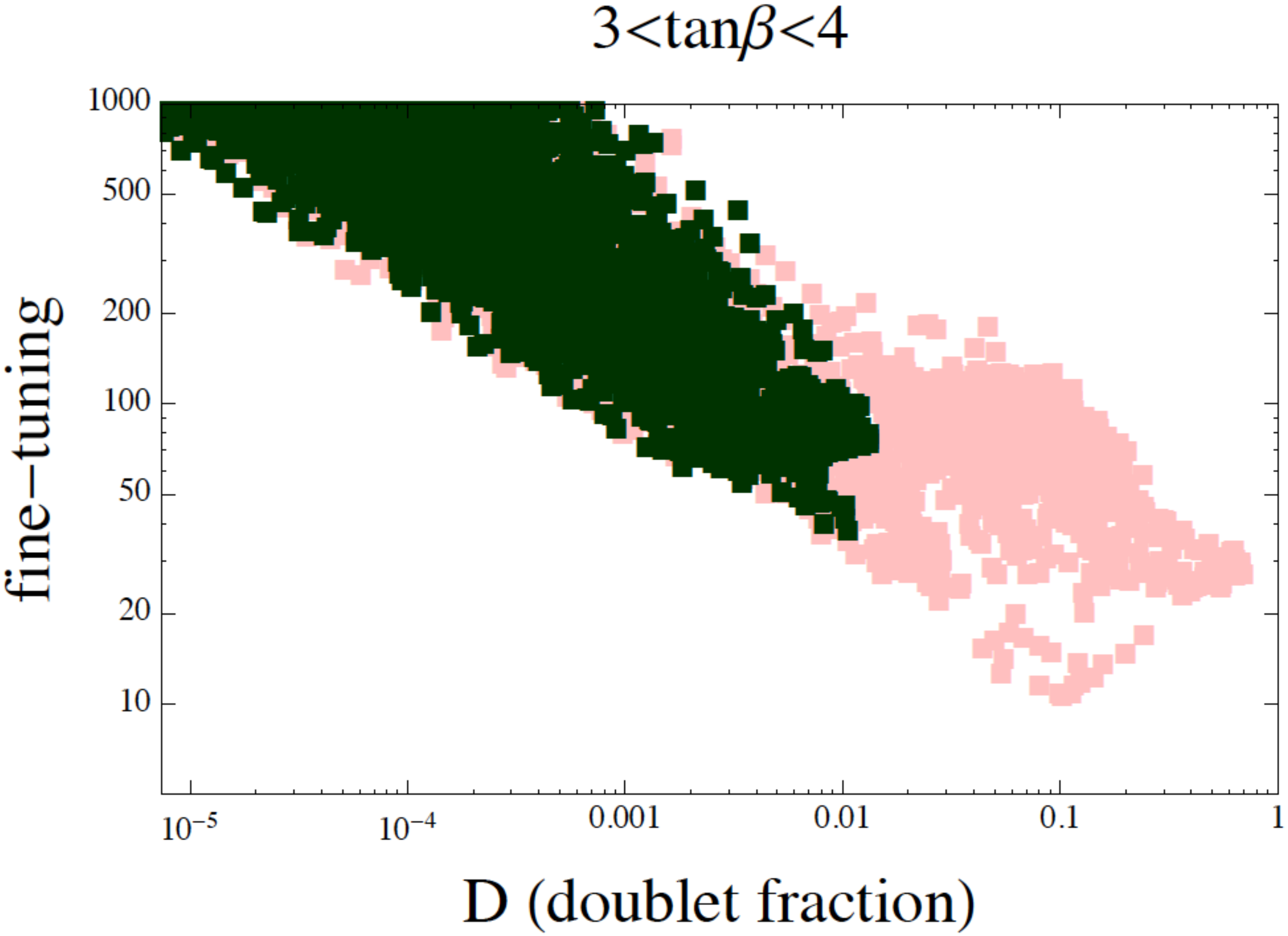}\\
\end{tabular}
\caption{Fine-tuning vs. the singlet fraction S (left) and the non-SM doublet fraction D (right) in the 126 GeV scalar, for $\lambda=2$ and various ranges of $\tan\beta$. Green points satisfy all constraints, while pink points satisfy all constraints except the LHC Higgs couplings fit. Points shown in yellow satisfy all constraints, but lie in the ``anomalous" region where loop corrections can be important for definitively establishing vacuum stability as well as consistency with LHC Higgs data.}
\label{fig:ftvsinglet}
\end{figure}

Fine-tuning and the composition of the 126 GeV Higgs are evaluated numerically for each point.
The correlations between fine-tuning $\Delta$ and the fractions of the singlet and the non-SM doublet in the 126 GeV scalar are shown in the scatter plots of Fig.~\ref{fig:ftvsinglet}. We discard points that fail any of the conditions listed in Section~\ref{sec:setup}, with the exception of the LHC bounds on Higgs couplings. Points that are excluded by the LHC Higgs fits at 95\%~c.l. but satisfy all other constraints are plotted in pink, while points that are allowed by these fits in addition to the other constraints are plotted in green or yellow. The yellow points are selected by the condition $|V_r| < 0.2 v^4$, indicating potential instability of the realistic vacuum. We checked that all such points with low fine-tuning (less than 100) are characterized by $s< 100$ GeV, and so belong to the anomalous region discussed at the end of the previous section.\footnote{The anomalous region disappears for larger $\tan\beta$, due to the LHC upper bound on the non-SM doublet fraction in the Higgs, $D$. For $\tan\beta\approx1$, $D$ is small regardless of the $H_v^0$ mass. For larger $\tan\beta$, suppressing $D$ requires raising the $H_v^0$ mass, which is not possible in the anomalous region.}
As we already explained, while this region is intriguing, its viability is questionable. Thus, in the rest of the discussion, we will ignore the yellow points in the scan and focus on the points falling in the bulk allowed regions of the parameter space, selected by $V_r \leq -0.2 v^4$ and shown in green, in which vacuum stability and all other constraints are robustly satisfied.

Two conclusions can be drawn from examining these points. First, {\it the least fine-tuned regions of $\lambda$-SUSY parameter space are already ruled out by the LHC Higgs rate measurements.} As discussed in the previous section, one would naively expect that about 15\% fine-tuning is required to obtain a 126 GeV Higgs for $\lambda=2$. Our scan contains many points with fine-tuning at this level, but none of them satisfy the current LHC bounds. The points that are still allowed have fine-tuning at the level of $2-3$\% at best, and most of the parameter space probed by our scans requires tuning at sub-per-cent level. Second, among the allowed points, there is a clear trend for fine-tuning to get higher as the singlet and doublet admixtures in the 126 GeV state decrease, {\it i.e.} as the Higgs becomes more SM-like. For the singlet fraction, this is precisely the trend that was observed for $\tan\beta=1$ in the previous section. Numerical scans confirm that this behavior persists for all $\tan\beta$, and that it applies to the non-SM doublet fraction (which is identically 0 for $\tan\beta=1$) as well.

\begin{figure}[t!]
\begin{center}
\includegraphics[width=0.6\linewidth]{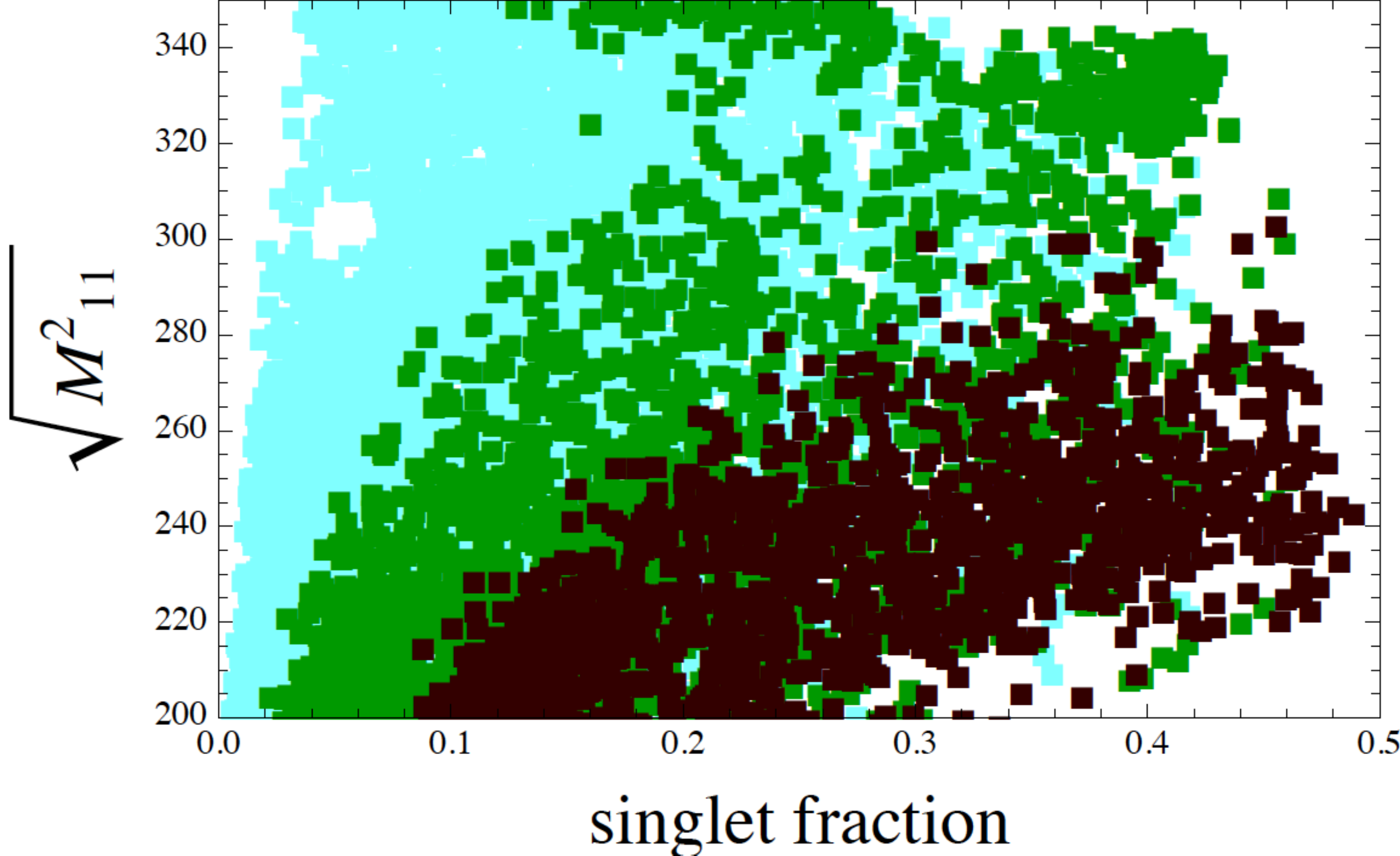}
\end{center}
\caption{Fine-tuning in the plane of SM-like Higgs mass before mixing and singlet fraction. Red, green, and cyan regions correspond to $\Delta<100, \,100<\Delta<300$, and $\Delta>300$ respectively. }
\label{fig:ls2bvssf}
\end{figure}

Another interesting feature in the scatter plots of Fig.~\ref{fig:ftvsinglet} is that points with larger values of $\tan\beta$ are systematically less fine-tuned. This is expected: the SM-like Higgs mass$^2$ before mixing is given by
\beq
{\cal M}_{11}^2 = \lambda^2v^2 \sin^22\beta + m_Z^2 \cos^2 2\beta.
\eeq{M11}
As discussed in the previous section, ${\cal M}_{11}^2$ is generally too large for $\lambda=2$, and mixing with other states is required to bring the Higgs mass down to the required $m_h=126$ GeV, resulting in fine-tuning. Since ${\cal M}_{11}^2$ decreases with increasing $\tan\beta$, the required cancellation is milder. To further illustrate this point, Fig.~\ref{fig:ls2bvssf} shows fine-tuning in the plane of $\sqrt{{\cal M}_{11}^2}$ and the singlet fraction. Lower fine-tuning is clearly correlated with both a lower $\sqrt{{\cal M}_{11}^2}$ and a larger singlet fraction. Note however that $\tan\beta$ cannot be increased beyond $4$ or so, due to precision electroweak constraints, so only a modest improvement in fine-tuning can be achieved.

\begin{figure}[t!]
\centering
\begin{tabular}{cc}
\includegraphics[width=3in,height=2in]{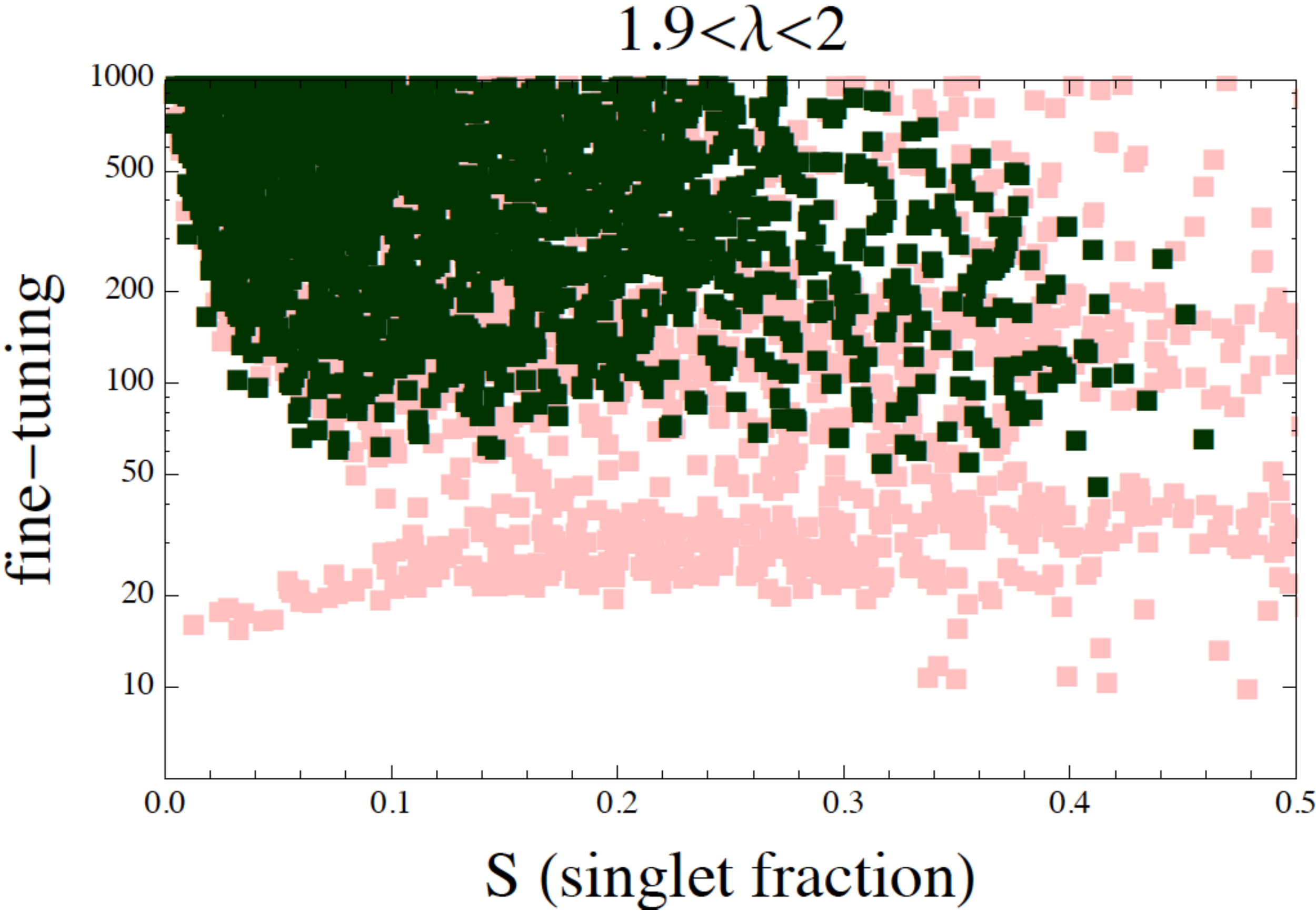}&
\includegraphics[width=3in,height=2in]{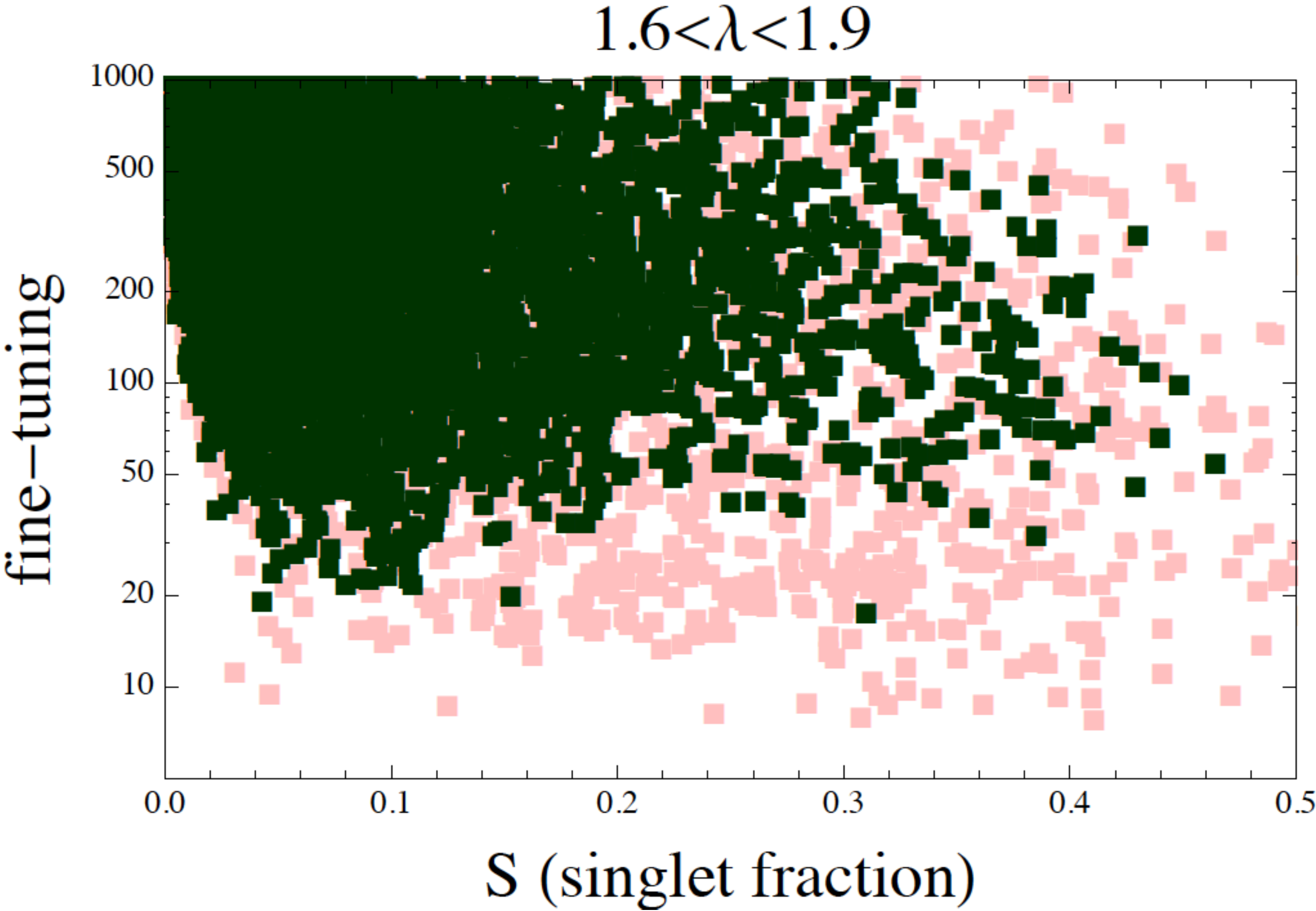}\\
\\
\includegraphics[width=3in,height=2in]{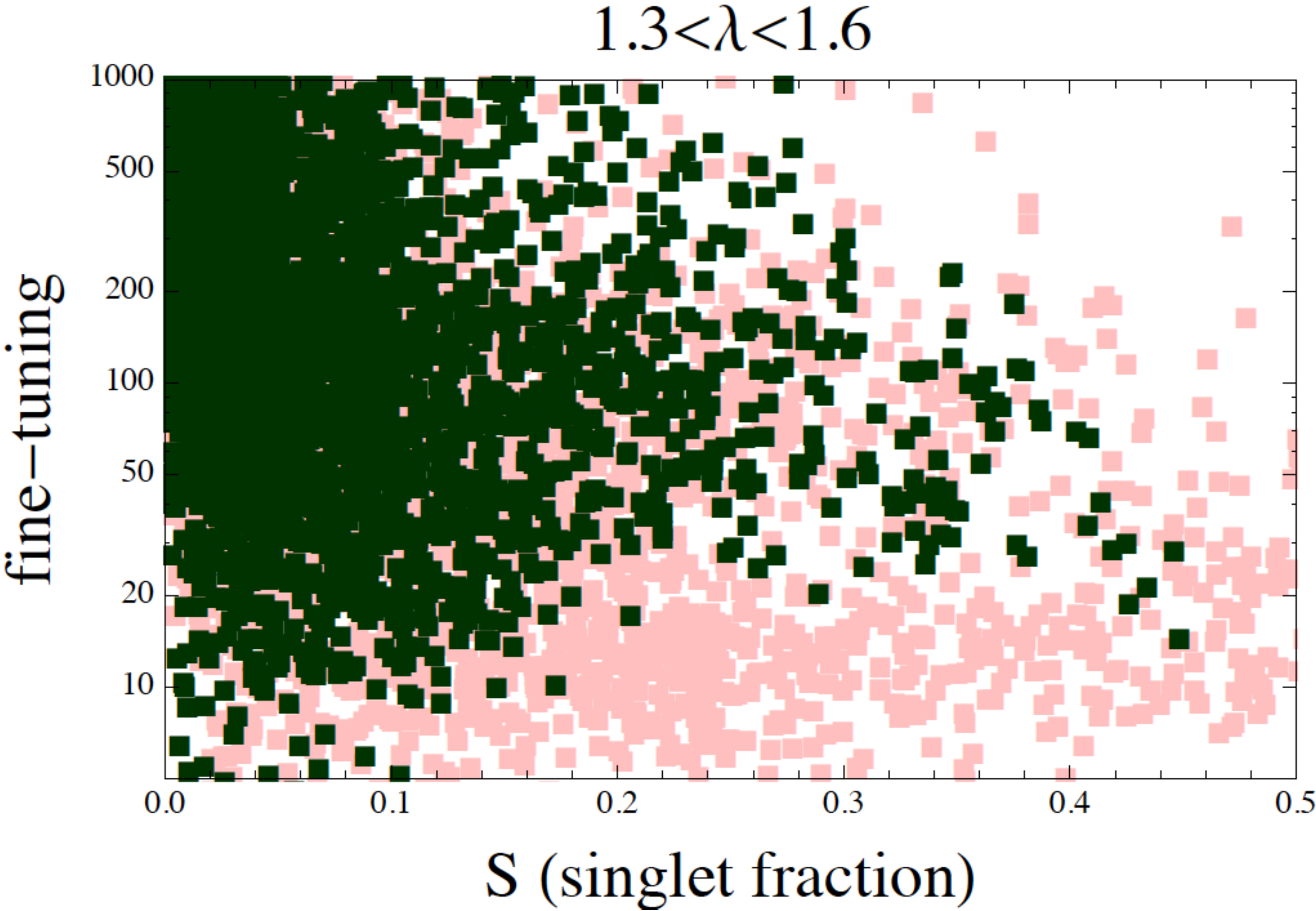}&
\includegraphics[width=3in,height=2in]{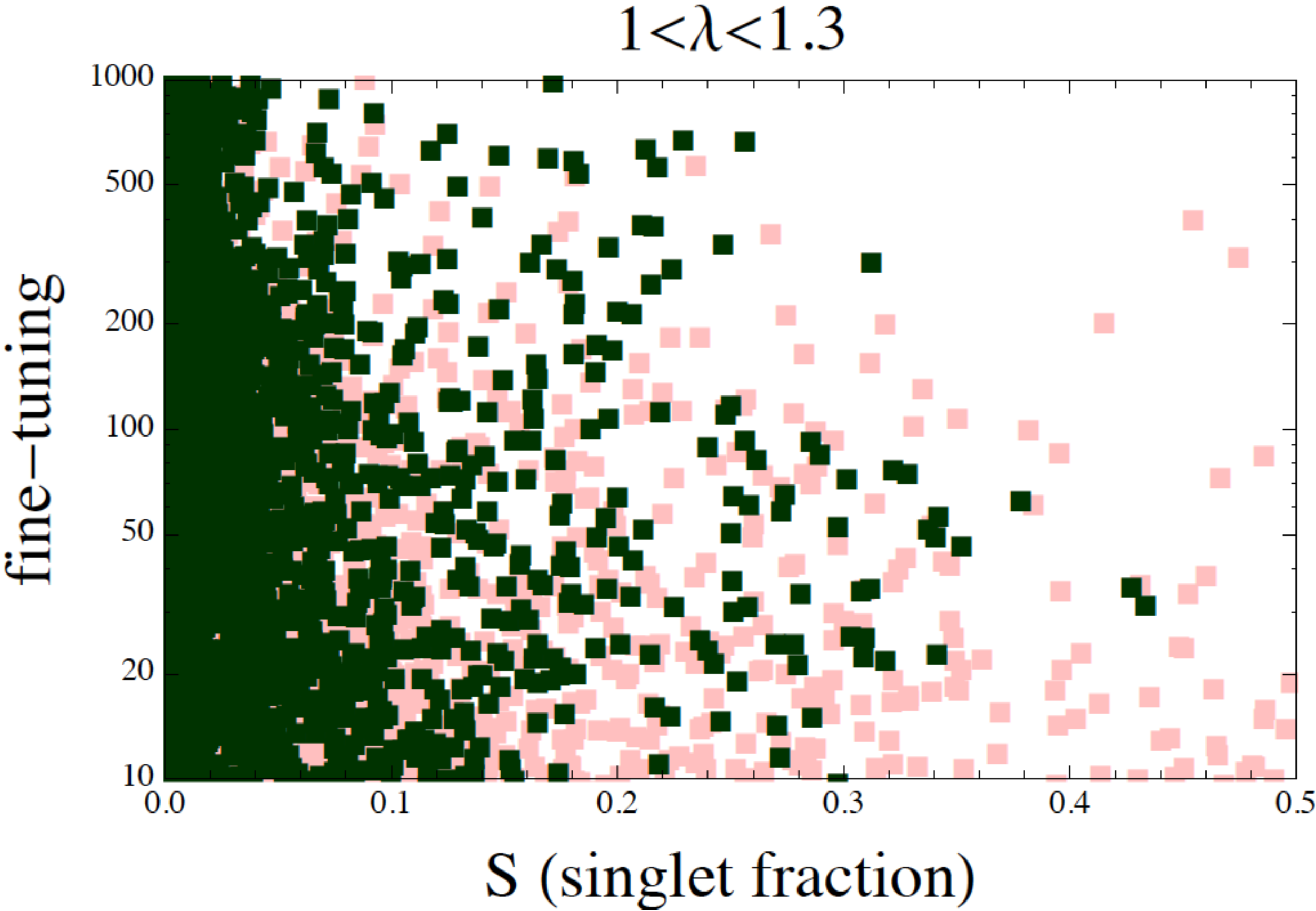}\\
\end{tabular}
\caption{Correlation of the Higgs singlet fraction and fine-tuning, for four ranges of the $\lambda$ parameter. Color code is the same as in Fig.~\ref{fig:ftvsinglet}, and there with $1<\tan\beta<4$. Anomalous points have been removed from the scans.}
\label{fig:lambdascan}
\end{figure}

It is also instructive to study the behavior of fine-tuning as $\lambda$ is varied. To do this, we repeated the scans, this time including $\lambda$ among the scanned variables. The results are shown if Fig.~\ref{fig:lambdascan}. The overall level of fine-tuning is clearly lower for lower $\lambda$. This trend has the same origin as the trend for lower fine-tuning at larger $\tan\beta$: reducing $\lambda$ also reduces the F-term contribution to ${\cal M}_{11}^2$, making it easier to obtain a 126 GeV eigenstate. Of course, it should be emphasized that this statement only applies to the tree-level fine-tuning measured by $\Delta$. As stressed in the Introduction, the fine-tuning associated with top sector loop corrections scales as $\sim\,(g/\lambda)^2$, and thus {\it increases} when $\lambda$ is decreased. The tendency of these two effects to move in opposite directions when $\lambda$ is varied has already been noted in Ref.~\cite{Gherghetta:2012gb}. The main features noted above for $\lambda=2$, the negative correlation of fine-tuning with the singlet fraction and the fact that the most natural part of the parameter space is ruled out by the LHC Higgs data, persist for $\lambda\gsim1.6$, but are lost at lower $\lambda$ where points with essentially no (tree-level) fine-tuning and small singlet fractions can be found. It should also be noted that as $\lambda$ decreases, the value of S for which fine-tuning is minimized also decreases; this simply reflects that as the tree-level mass decreases, a smaller amount of mixing with the singlet is required to reach 126 GeV.

\begin{figure}[t!]
\centering
\begin{tabular}{cc}
\includegraphics[width=3in,height=2in]{beforecut2}&
\includegraphics[width=3in,height=2in]{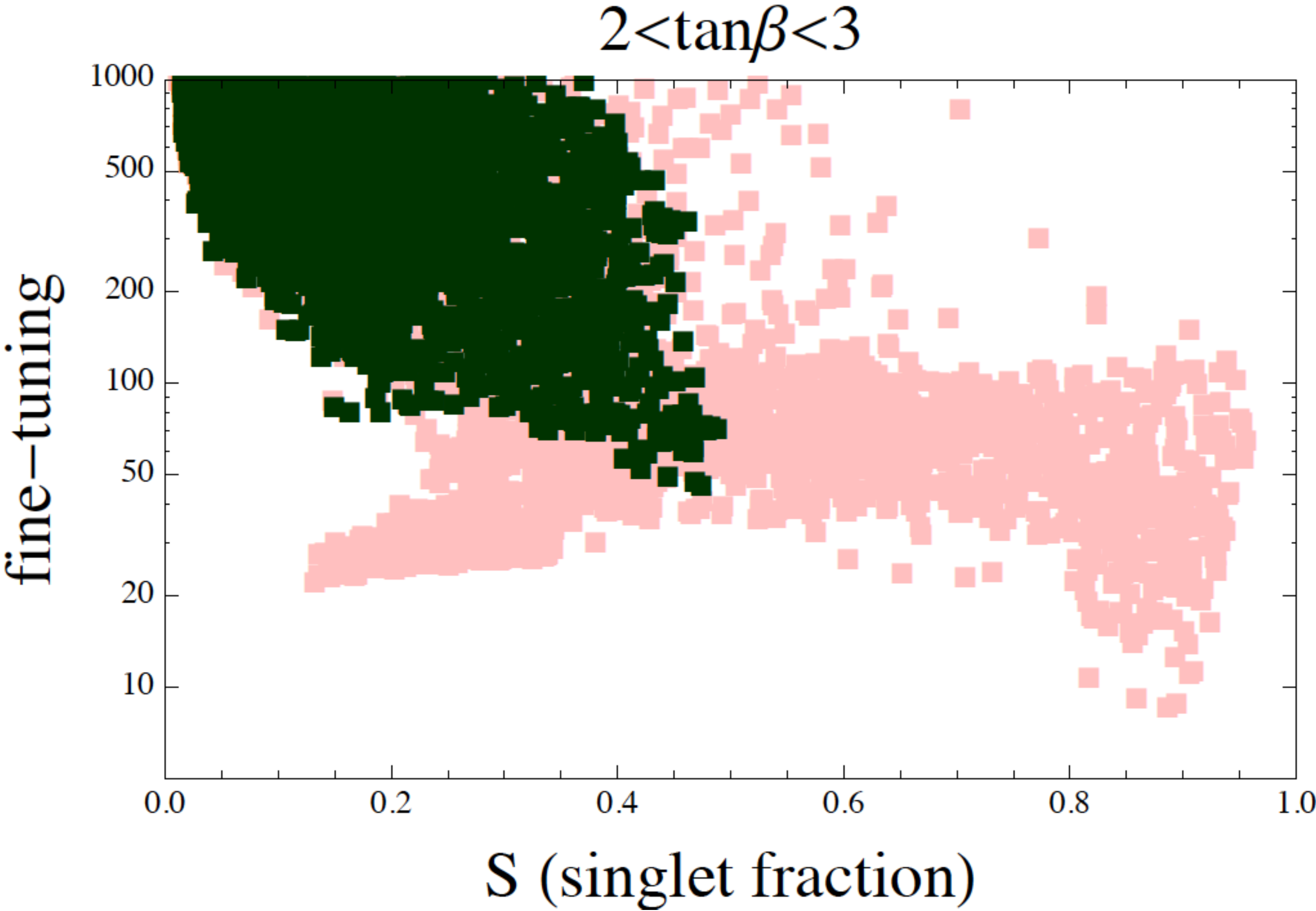}\\
\end{tabular}
\caption{Correlation of the Higgs singlet fraction and fine-tuning, for $\lambda=2$ and $\tan\beta\in [2, 3]$. Left: tree-level analysis, same as Fig.~\ref{fig:ftvsinglet}. Right: same analysis, but including the top sector one-loop correction to the up-type Higgs mass. Stop masses of 1 TeV and no mixing are assumed. Color code is the same as in Fig.~\ref{fig:ftvsinglet}. }
\label{fig:top_loop}
\end{figure}

Finally, to study the importance of loop corrections, we repeated the scans including the top/stop one-loop contribution to the CP-even Higgs mass$^2$ matrix with a stop mass of 1 TeV and no mixing. We find that all qualitative features discussed above remain unchanged, while the overall level of fine-tuning is slightly higher. An example of this is shown in Fig.~\ref{fig:top_loop}. This behavior is easily understood: the top/stop loops give an additional positive contribution to ${\cal M}_{11}^2$, requiring a stronger cancellation to obtain a 126 GeV eigenvalue (see, for example, Eq.~\leqn{mhapp}). Since the stop contribution is subdominant compared to the large tree-level entries in the mass matrix for large $\lambda$, the overall increase in fine-tuning is correspondingly small: in this case, less than a factor of 2 for 1 TeV stops. The fine-tuning is expected to increase with increasing stop mass. Again, we emphasize that $\Delta$ measures only the sensitivity to tree-level parameters, so that this fine-tuning is {\it in addition} to the well-known fine-tuning due to the sensitivity of the weak scale to the stop mass at loop level.

\section{Conclusions}
\label{sec:conc}

The NMSSM provides an attractive possibility to realize supersymmetry at the weak scale, consistent with the LHC discovery of a Higgs boson at 126 GeV. At the same time, in the $\lambda$-SUSY regime, it can also address the tension between naturalness and non-observation of superpartners at the LHC. The lower bounds on stop masses have now been pushed beyond the $\sim 500$ GeV bound where the minimal supersymmetric model can be completely natural, at least for simple spectra with light LSPs. In $\lambda$-SUSY, the fine-tuning is reduced by a factor $\sim(g/\lambda)^2$ compared to the MSSM with the same stop mass, so that the current bounds are not necessarily in conflict with naturalness. This prompts a serious consideration of this model.

Unfortunately, with the discovery of a SM-like Higgs at 126 GeV, large values of $\lambda$ introduce an additional fine-tuning, an anomalous sensitivity of the Higgs mass to the tree-level potential parameters. This is due to the simple fact that the tree-level potential, for large $\lambda$ and moderate $\tan\beta$, produces a doublet Higgs mass well in excess of the observed 126 GeV, which then needs to be cancelled by mixing the doublet and singlet Higgs states. In this paper, we showed that this tree-level fine-tuning is even stronger than naively expected, when the LHC bounds on the Higgs couplings are taken into account. The measured Higgs couplings are consistent with SM predictions, placing tight bounds on the possible mixing with a singlet or a non-SM doublet Higgs states. In $\lambda$-SUSY, such mixing is generic. Once the constraints needed to ensure viability of the model (such as absence of tachyons and stability of the EWSB vacuum with respect to tunneling into lower-lying vacua) are imposed, the mixings can only be made small at the expense of an additional fine-tuning. The current LHC bounds already imply tree-level fine-tuning at the level of  $2-3$\% at best for $\lambda=2.0$. (For smaller $\lambda$, the tree-level fine-tuning can be reduced, but only at the expense of re-introducing fine-tuning with respect to the stop mass at one-loop level.) Moreover, a strong negative correlation exists between deviations of the Higgs couplings from SM and the required fine-tuning. Any further improvement in the Higgs coupling measurements would either yield a discovery of a deviation from the SM, or rule out the most natural remaining parameter regions of $\lambda$-SUSY. This adds to the already long list of motivations to measure these couplings as precisely as possible.

Interestingly, we found a small part of the model parameter space, the ``anomalous region", where relatively small fine-tuning (of order 10\% or less) seems to be achievable. However, viability of this region can only be established definitively if loop-level corrections, both in the scalar potential and in the Higgs couplings to photons and gluons, are included. We leave such an analysis for future work.

\vskip0.8cm
\noindent{\large \bf Acknowledgments}
\vskip0.3cm

We would like to thank Roberto Franceschini, Benedict von Harling, Filippo Sala and Andrea Tesi for useful discussions. MF would like to thank the Galileo Galilei Institute in Florence and the Aspen Center for Physics for their hospitality. MP would like to acknowledge the hospitality of KITP at Santa Barbara, where part of this work was completed. Similarly, BS would like to thank CETUP* (Center for Theoretical Underground Physics and Related Areas), supported by the US Department of Energy under Grant No. DE-SC0010137 and by the US National Science Foundation under Grant No. PHY-1342611, for its hospitality and partial support during the 2013 Summer Program. This research is supported by the U.S. National Science Foundation through grant PHY-0757868 and CAREER grant PHY-0844667. BS is also supported by DoE grant DE-SC0007859.


\begin{thebibliography}{99}

\bibitem{Barbieri:2000gf}
  R.~Barbieri and A.~Strumia,
{\it ``The 'LEP paradox',''}
  [\hhref{hep-ph/0007265}].

\bibitem{Ellwanger:2009dp}
  U.~Ellwanger, C.~Hugonie and A.~M.~Teixeira,
{\it ``The Next-to-Minimal Supersymmetric Standard Model,''}
  Phys.\ Rept.\  {\bf 496}, 1 (2010)
  [\hhref{0910.1785} [hep-ph]].

\bibitem{Hardy:2012ef}
  E.~Hardy, J.~March-Russell and J.~Unwin,
{\it ``Precision Unification in $\lambda$ SUSY with a 125 GeV Higgs,''}
  JHEP {\bf 1210}, 072 (2012)
  [arXiv:1207.1435 [hep-ph]].

\bibitem{Barbieri:2006bg}
  R.~Barbieri, L.~J.~Hall, Y.~Nomura and V.~S.~Rychkov,
{\it ``Supersymmetry without a Light Higgs Boson,''}
  Phys.\ Rev.\ D {\bf 75}, 035007 (2007)
  [\hhref{hep-ph/0607332}].

\bibitem{Hall:2011aa}
  L.~J.~Hall, D.~Pinner and J.~T.~Ruderman,
{\it ``A Natural SUSY Higgs Near 126 GeV,''}
  JHEP {\bf 1204}, 131 (2012)
  [\hhref{1112.2703} [hep-ph]].

\bibitem{Perelstein:2012qg}
  M.~Perelstein and B.~Shakya,
{\it ``XENON100 Implications for Naturalness in the MSSM, NMSSM and lambda-SUSY,''}
  [\hhref{1208.0833} [hep-ph]].

\bibitem{Agashe:2012zq}
  K.~Agashe, Y.~Cui and R.~Franceschini,
 {\it ``Natural Islands for a 125 GeV Higgs in the scale-invariant NMSSM,''}
  JHEP {\bf 1302}, 031 (2013)
  [\hhref{1209.2115} [hep-ph]].

\bibitem{Gherghetta:2012gb}
  T.~Gherghetta, B.~von Harling, A.~D.~Medina and M.~A.~Schmidt,
 {\it ``The Scale-Invariant NMSSM and the 126 GeV Higgs Boson,''}
  JHEP {\bf 1302}, 032 (2013)
  [\hhref{1212.5243} [hep-ph]].

\bibitem{King:2012tr}
  S.~F.~King, M.~Mühlleitner, R.~Nevzorov and K.~Walz,
  {\it``Natural NMSSM Higgs Bosons,''}
  Nucl.\ Phys.\ B {\bf 870}, 323 (2013)
  [\hhref{1211.5074} [hep-ph]].

\bibitem{Gupta:2012fy}
  R.~S.~Gupta, M.~Montull and F.~Riva,
  {\it``SUSY Faces its Higgs Couplings,''}
  JHEP {\bf 1304}, 132 (2013)
  [\hhref{1212.5240} [hep-ph]].

\bibitem{Cheung:2013bn}
  C.~Cheung, S.~D.~McDermott and K.~M.~Zurek,
  {\it``Inspecting the Higgs for New Weakly Interacting Particles,''}
  JHEP {\bf 1304}, 074 (2013)
  [\hhref{1302.0314} [hep-ph]].

\bibitem{Cheng:2013fma}
  T.~Cheng, J.~Li, T.~Li and Q.~-S.~Yan,
  {\it``Natural NMSSM confronting with the LHC7-8,''}
  [\hhref{1304.3182} [hep-ph]].

\bibitem{Barbieri:2013hxa}
  R.~Barbieri, D.~Buttazzo, K.~Kannike, F.~Sala and A.~Tesi,
  {\it``Exploring the Higgs sector of a most natural NMSSM,''}
  Phys.\ Rev.\ D {\bf 87}, 115018 (2013)
  [\hhref{1304.3670} [hep-ph]].

\bibitem{Cao:2012yn}
  J.~Cao, Z.~Heng, J.~M.~Yang and J.~Zhu,
{\it ``Status of low energy SUSY models confronted with the LHC 125 GeV Higgs data,''}
  JHEP {\bf 1210}, 079 (2012)
  [arXiv:1207.3698 [hep-ph]].

\bibitem{Peskin:2012we}
  M.~E.~Peskin,
{\it ``Comparison of LHC and ILC Capabilities for Higgs Boson Coupling Measurements,''}
  [\hhref{1207.2516} [hep-ph]].

\bibitem{Baer:2013cma}
  H.~Baer~
  {\it et al.},
  {\it ``The International Linear Collider Technical Design Report - Volume 2: Physics,''}
  [\hhref{1306.6352} [hep-ph]].

\bibitem{Romao:1986jy}
  J.~C.~Romao,
{\it ``Spontaneous {CP} Violation in {SUSY} Models: A No Go Theorem,''}
  Phys.\ Lett.\ B {\bf 173} (1986) 309.


\bibitem{ATLASbb}
  [ATLAS Collaboration],
  ATLAS-CONF-2012-161.

\bibitem{ATLAStautau}
  G.~Aad {\it et al.}  [ATLAS Collaboration],
  {\it``Search for the Standard Model Higgs boson in the $H$ to $\tau^{+} \tau^{-}$ decay mode in $\sqrt{s}=7$ TeV $pp$ collisions with ATLAS,''}
  JHEP {\bf 1209}, 070 (2012)
  [\hhref{1206.5971} [hep-ex]];
  ATLAS-CONF-2012-160.

\bibitem{ATLASww}
  G.~Aad {\it et al.}  [ATLAS Collaboration],
  {\it``Search for the Standard Model Higgs boson in the $H \to$ WW(*) $\to \ell \nu \ell \nu$ decay mode with 4.7 /fb of ATLAS data at $\sqrt{s}=7$ TeV,''}
  Phys.\ Lett.\ B {\bf 716}, 62 (2012)
  [\hhref{1206.0756} [hep-ex]]; ATLAS-CONF-2013-030.

\bibitem{ATLASzz}
  G.~Aad {\it et al.}  [ATLAS Collaboration],
  {\it``Search for the Standard Model Higgs boson in the decay channel $H \to$ ZZ(*) $\to 4 \ell$ with 4.8 fb-1 of $pp$ collision data at $\sqrt{s}=7$ TeV with ATLAS,''}
  Phys.\ Lett.\ B {\bf 710}, 383 (2012)
  [\hhref{1202.1415} [hep-ex]];
ATLAS-CONF-2013-013.

\bibitem{ATLASgaga}
  G.~Aad {\it et al.}  [ATLAS Collaboration],
  {\it``Search for the Standard Model Higgs boson in the diphoton decay channel with 4.9 fb$^{-1}$ of $pp$ collisions at $\sqrt{s}=7$ TeV with ATLAS,''}
  Phys.\ Rev.\ Lett.\  {\bf 108}, 111803 (2012)
  [\hhref{1202.1414} [hep-ex]].

 \bibitem{ATLASgaga1212}
 ATLAS Collaboration,   ATLAS-CONF-2013-012.

  \bibitem{CMSbb}
    S.~Chatrchyan {\it et al.}  [CMS Collaboration],
  {\it``Search for the standard model Higgs boson decaying to bottom quarks in pp collisions at sqrt(s)=7 TeV,''}
  Phys.\ Lett.\ B {\bf 710}, 284 (2012)
  [\hhref{1202.4195} [hep-ex]].

\bibitem{CMStautau}
CMS Collaboration,  CMS-PAS-HIG-12-043; CMS-PAS-HIG-13-004.

\bibitem{CMSww}
CMS Collaboration,  CMS-HIG-12-042;  CMS-HIG-13-003; CMS-HIG-13-009.

 \bibitem{CMSzz}
  CMS Collaboration,
  CMS-HIG-12-023;
    CMS-HIG-13-002.

\bibitem{CMSgaga}
  CMS Collaboration,
  {\it``Evidence for a new state decaying into two photons in the search for the standard model Higgs boson in pp collisions,''}
  CMS-PAS-HIG-12-015; CMS-PAS-HIG-13-001.

\bibitem{Tevatron}
  T.~Aaltonen {\it et al.}  [CDF and D0 Collaborations],
  {\it``Higgs Boson Studies at the Tevatron,''}
  [\hhref{1303.6346} [hep-ex]].


\bibitem{Ellwanger:2005fh}
  U.~Ellwanger and C.~Hugonie,
{\it ``Yukawa induced radiative corrections to the lightest Higgs boson mass in the NMSSM,''}
  Phys.\ Lett.\ B {\bf 623}, 93 (2005)
  [\hhref{hep-ph/0504269}].

\bibitem{Degrassi:2009yq}
  G.~Degrassi and P.~Slavich,
{\it ``On the radiative corrections to the neutral Higgs boson masses in the NMSSM,''}
  Nucl.\ Phys.\ B {\bf 825}, 119 (2010)
  [\hhref{0907.4682} [hep-ph]].

\bibitem{Staub:2010ty}
  F.~Staub, W.~Porod and B.~Herrmann,
 {\it ``The Electroweak sector of the NMSSM at the one-loop level,''}
  JHEP {\bf 1010}, 040 (2010)
  [arXiv:1007.4049 [hep-ph]].

  \bibitem{King:1995vk}
  S.~F.~King and P.~L.~White,
  {\it``Resolving the constrained minimal and next-to-minimal supersymmetric standard models,''}
  Phys.\ Rev.\ D {\bf 52}, 4183 (1995)
  [\hhref{hep-ph/9505326}].

  \bibitem{Masip:1998jc}
  M.~Masip, R.~Munoz-Tapia and A.~Pomarol,
  {\it```Limits on the mass of the lightest Higgs in supersymmetric models,''}
  Phys.\ Rev.\ D {\bf 57}, R5340 (1998)
  [\hhref{hep-ph/9801437}].

\bibitem{Franceschini:2010qz}
  R.~Franceschini and S.~Gori,
  {\it``Solving the $\mu$ problem with a heavy Higgs boson,''}
  JHEP {\bf 1105}, 084 (2011)
  [\hhref{1005.1070} [hep-ph]].

\bibitem{Belanger:2012tt}
  G.~Belanger, U.~Ellwanger, J.~F.~Gunion, Y.~Jiang, S.~Kraml and J.~H.~Schwarz,
  {\it``Higgs Bosons at 98 and 125 GeV at LEP and the LHC,''}
  JHEP {\bf 1301}, 069 (2013)
  [\hhref{1210.1976} [hep-ph]].

\bibitem{Kang:2013rj}
  Z.~Kang, J.~Li, T.~Li, D.~Liu and J.~Shu,
  {\it``Probing the CP-even Higgs Sector via $H_3\to H_2H_1$ in the Natural NMSSM,''}
  Phys.\ Rev.\ D {\bf 88}, 015006 (2013)
  [\hhref{1301.0453} [hep-ph]].

\bibitem{Badziak:2013bda}
  M.~Badziak, M.~Olechowski and S.~Pokorski,
  {\it``New Regions in the NMSSM with a 125 GeV Higgs,''}
  JHEP {\bf 1306}, 043 (2013)
  [\hhref{1304.5437} [hep-ph]].

\bibitem{Barbieri:2013nka}
  R.~Barbieri, D.~Buttazzo, K.~Kannike, F.~Sala and A.~Tesi,
  {\it``One or more Higgs bosons?,''}
  [\hhref{1307.4937} [hep-ph]].

\bibitem{Cao:2013gba}
  J.~Cao, F.~Ding, C.~Han, J.~M.~Yang and J.~Zhu,
{\it ``A light Higgs scalar in the NMSSM confronted with the latest LHC Higgs data,''}
  arXiv:1309.4939 [hep-ph].


\bibitem{CMS:gya}
  [CMS Collaboration],
  {\it ``Higgs to tau tau (MSSM) (HCP),''}
  CMS-PAS-HIG-12-050.


\bibitem{Beringer:1900zz}
  J.~Beringer {\it et al.}  [Particle Data Group Collaboration],
  {\it``Review of Particle Physics (RPP),''}
  Phys.\ Rev.\ D {\bf 86}, 010001 (2012).



  \bibitem{PDG}
  J. Beringer et al. (Particle Data Group), Phys. Rev. D86, 010001 (2012)

  \bibitem{Hooper:2002nq}
  D.~Hooper and T.~Plehn,
  {\it``Supersymmetric dark matter: How light can the LSP be?,''}
  Phys.\ Lett.\ B {\bf 562}, 18 (2003)
  [hep-ph/0212226].

\bibitem{Kanehata:2011ei}
  Y.~Kanehata, T.~Kobayashi, Y.~Konishi, O.~Seto and T.~Shimomura,
  {\it``Constraints from Unrealistic Vacua in the Next-to-Minimal Supersymmetric Standard Model,''}
  Prog.\ Theor.\ Phys.\  {\bf 126}, 1051 (2011)
  [\hhref{1103.5109} [hep-ph]].

\bibitem{Kobayashi:2012xv}
  T.~Kobayashi, T.~Shimomura and T.~Takahashi,
  {\it``Constraining the Higgs sector from False Vacua in the Next-to-Minimal Supersymmetric Standard Model,''}
  Phys.\ Rev.\ D {\bf 86}, 015029 (2012)
  [\hhref{1203.4328} [hep-ph]].

\bibitem{Perelstein:2007nx}
  M.~Perelstein and C.~Spethmann,
 {\it ``A Collider signature of the supersymmetric golden region,''}
  JHEP {\bf 0704}, 070 (2007)
  [\hhref{hep-ph/0702038}].

\bibitem{Baer:2012up}
  H.~Baer, V.~Barger, P.~Huang, A.~Mustafayev and X.~Tata,
{\it ``Radiative natural SUSY with a 125 GeV Higgs boson,''}
  Phys.\ Rev.\ Lett.\  {\bf 109}, 161802 (2012)
  [\hhref{1207.3343} [hep-ph]].

\bibitem{Baer:2012mv}
  H.~Baer, V.~Barger, P.~Huang, D.~Mickelson, A.~Mustafayev and X.~Tata,
 {\it ``Post-LHC7 fine-tuning in the mSUGRA/CMSSM model with a 125 GeV Higgs boson,''}
  Phys.\ Rev.\ D {\bf 87}, no. 3, 035017 (2013)
  [\hhref{1210.3019} [hep-ph]].

\bibitem{Baer:2013gva}
  H.~Baer, V.~Barger and D.~Mickelson,
 {\it ``How conventional measures overestimate electroweak fine-tuning in supersymmetric theory,''}
  [\hhref{1309.2984} [hep-ph]].


 \bibitem{ftmeasure}
  U.~Ellwanger, G.~Espitalier-Noel and C.~Hugonie,
  {\it``Naturalness and fine tuning in the NMSSM: Implications of Early LHC Results,''}
  JHEP {\bf 1109}, 105 (2011)
  [\hhref{1107.2472} [hep-ph]].

\bibitem{SchmidtHoberg:2012yy}
  K.~Schmidt-Hoberg and F.~Staub,
 {\it ``Enhanced $h\rightarrow \gamma \gamma$ rate in MSSM singlet extensions,''}
  JHEP {\bf 1210}, 195 (2012)
  [arXiv:1208.1683 [hep-ph]].

\bibitem{Choi:2012he}
  K.~Choi, S.~H.~Im, K.~S.~Jeong and M.~Yamaguchi,
  {\it``Higgs mixing and diphoton rate enhancement in NMSSM models,''}
  JHEP {\bf 1302}, 090 (2013)
  [\hhref{1211.0875} [hep-ph]].

\end{thebibliography}
\end{document}